\documentclass[12pt]{article}
\usepackage[utf8]{inputenc}
\usepackage{amsfonts,amsmath,amssymb,mathtools}
\usepackage[left=2cm,right=2cm,top=2cm,bottom=2cm,bindingoffset=0cm]{geometry}
\usepackage{graphicx}
\usepackage{cite}
\usepackage{authblk}
\usepackage{slashed}
\usepackage{braket}
\usepackage{xcolor}
\usepackage{hyperref}

\newcommand\bem{\begin{pmatrix}}
\newcommand\eem{\end{pmatrix}}
\newcommand\beq{\begin{equation}}
\newcommand\eeq{\end{equation}}
\newcommand\beqs{\begin{equation*}}
\newcommand\eeqs{\end{equation*}}

\newcommand{\pd}{\partial}
\newcommand{\sgn}{\text{sgn}}
\newcommand{\mO}{\mathcal{O}}
\newcommand{\mT}{\mathcal{T}}
\newcommand{\const}{\text{const}}

\numberwithin{equation}{section}

\hypersetup{
    colorlinks=true,
    linkcolor=black,
    citecolor=blue,   
    urlcolor=blue,
}

\title{\bf Dynamical Casimir effect in nonlinear vibrating cavities}
\author[1,2,3]{Lianna~A.~Akopyan\thanks{\href{mailto:lianna.akopyan@phystech.edu}{lianna.akopyan@phystech.edu}}}
\author[1,2]{Dmitrii~A.~Trunin\thanks{\href{mailto:dmitriy.trunin@phystech.edu}{dmitriy.trunin@phystech.edu}}}
\affil[1]{Moscow Institute of Physics and Technology, 141700, Institutskii per., 9, Dolgoprudny, Russia}
\affil[2]{Institute for Theoretical and Experimental Physics, 117218, B. Cheremushkinskaya, 25, Moscow, Russia}
\affil[3]{Russian Quantum Center, 121205, Bolshoy Boulevard, 30, bld. 1, Moscow, Russia}
\date{\today}

\begin{document}

\maketitle

\begin{abstract}
Nonlinear terms in the equations of motion can induce secularly growing loop corrections to correlation functions. Recently such corrections were shown to affect the particle production by a nonuniformly moving ideal mirror. We extend this conclusion to the cases of ideal vibrating cavity and single semitransparent mirror. These models provide natural IR and UV scales and allow a more accurate study of the loop behavior. In both cases we confirm that two-loop correction to the Keldysh propagator quadratically grows with time. This growth indicates a breakdown of the semiclassical approximation and emphasizes that bulk nonlinearities in the dynamical Casimir effect cannot be neglected for large evolution times.
\end{abstract}

\newpage

\tableofcontents

\section{Introduction}
\label{sec:intro}

A nonuniformly accelerated mirror is known to create particles due to fluctuations of the ubiquitous quantum fields. This effect, which is often referred to as the dynamical Casimir effect (DCE), was theoretically predicted 50 years ago by G.~T.~Moore who considered a simplified model of one-dimensional ideal cavity~\cite{Moore}. Subsequently it was also extended to a single-mirror case~\cite{DeWitt, Davies:1976, Davies:1977, Birrell}. Similarly to the Hawking~\cite{Hawking} and Unruh~\cite{Unruh, Fulling, Davies:1974} effects, DCE has no classical counterparts and illustrates the most fundamental features of quantum field theory.

These features are most clearly manifested in a simplified two-dimensional model which admits relatively simple analytical solutions. This is the model of a free massless scalar field with time-dependent Dirichlet boundary conditions:
\beq \label{eq:simple}
(\pd_t^2 - \pd_x^2) \phi(t,x) = 0, \quad \phi\left[ t, L(t) \right] = \phi\left[ t, R(t) \right] = 0, \eeq
where functions $L(t)$ and $R(t)$ determine the positions of two perfectly reflective mirrors at the moment $t$. Most of the papers on the DCE (including seminal ones~\cite{Moore, DeWitt, Davies:1976, Davies:1977}) are devoted to this simplified model. In particular, the creation of scalar particles by such boundary conditions has been extensively investigated by different approaches to the mode decomposition of the quantized field~\cite{Cole, Meplan, Li, Law:1994, Law:1995, Dodonov:1996, Dalvit:1997, Dalvit:1998, Kim} and calculation of the effective action~\cite{Fosco, Nagatani, Fosco:2017}. Generalizations of the scalar DCE with imperfect mirrors were considered in~\cite{Lambrecht, Barton, Nicolaevici, Obadia}. Moreover, the simplified model~\eqref{eq:simple} can be implemented using superconducting circuits~\cite{Johansson:2009, Johansson:2010, Nation}; this idea led to the first experimental observations of the DCE~\cite{Wilson, Lahteenmaki}. A comprehensive review of the current theoretical and experimental status of the DCE can be found in~\cite{Dodonov:2010, Dodonov:2020}.

We emphasize that the existing theoretical and experimental studies of the DCE are mainly limited to semiclassical (tree-level) effects, i.e. they focus on the linearized eq.~\eqref{eq:simple} or its analogs. At the same time, in nonstationary quantum field theory semiclassical approximation often fails to provide the complete answer, because quantum loop corrections substantially affect the state of the system. Furthermore, loop corrections to the tree-level correlation functions, quantum averages and stress-energy flux indefinitely grow with time. Therefore, at large evolution times such corrections are significant even if nonlinear terms in the classical equations of motion seem negligible. For example, loop corrections considerably affect the particle creation processes in an expanding universe~\cite{Krotov, Akhmedov, Burda, Polyakov, Akhmedov:dS, Popov, Pavlenko, Moschella}, strong electric~\cite{Musaev, Akhmedov:Et, Akhmedov:Ex}, scalar~\cite{Diatlyk-1, Trunin-1, Diatlyk-2} and gravitational~\cite{Akhmedov:H} fields.

Recently loop corrections were also shown to play an important role in the DCE~\cite{Alexeev}. Namely, it was shown that in a simplified two-dimensional model of the DCE, which describes a massless real scalar field on a single ideal mirror background, quantum loop corrections to the Keldysh propagator indefinitely grow with the evolution time. This means that at large evolution times the semiclassical approximation is not applicable. Furthermore, the Keldysh propagator is closely related to the stress-energy flux and the state of the quantum field, so loop corrections evidently affect the particle creation in the DCE.

In this paper we extend the results of~\cite{Alexeev} to two more realistic models. First, we examine the case of a vibrating cavity bounded by two perfectly reflecting mirrors. On the one hand, this model provides a natural IR scale and allows a more accurate study of the secular growth. On the other hand, it has more experimental significance than the single-mirror model. Second, we discuss the case of a single nonideal (semitransparent) mirror which is well-defined in the UV region.

In both cases we calculate two-loop corrections to the Keldysh propagator of a two-dimensional massless real scalar field with quartic ($\lambda \phi^4$) self-interaction. This calculation essentially reduces to the calculation of the energy level density and anomalous quantum average which are parts of the Keldysh propagator. We find that in both cases loop corrections quadratically grow with time. Also we show that this growth is associated with the violation of the conformal invariance by the $\lambda \phi^4$ interaction term.

This paper is organized as follows. In Secs.~\ref{sec:SK} and~\ref{sec:picture} we briefly review the nonequilibrium Schwinger--Keldysh diagrammatic technique and discuss the physical picture underlying the scalar DCE. In the following sections we use this technique to calculate loop corrections to the quantum averages of a massless scalar field on various nonstationary backgrounds. In Sec.~\ref{sec:two} we consider the case of two nonuniformly moving ideal mirrors. We employ the geometrical method of constructing modes to simplify the calculation of loop integrals. Also we illustrate this calculation by examples of resonant cavity, synchronized and unsynchronized ``broken'' mirror trajectories. In Sec.~\ref{sec:non} we discuss the case of a single nonideal mirror. For simplicity we assume that the proper acceleration of the mirror is much smaller than the energy scale of semitransparency. Finally, we discuss the results and conclude in Sec.~\ref{sec:discussion}. In addition, we consider the stationary case in App.~\ref{sec:artifacts} and discuss miscellaneous calculation details in Apps.~\ref{sec:expl-modes} and~\ref{sec:loop_cor}.

\subsection{Schwinger--Keldysh diagrammatic technique}
\label{sec:SK}

In this paper we study quantum loop corrections to the scalar DCE which can be described by the following action (we set $\hbar = c = 1$):
\beq \label{eq:S}
S = \int d^2 x \left[ \frac{1}{2} \left( \pd_\mu \phi(t,x) \right)^2 - \frac{1}{2} V(t,x) \phi^2(t,x) - \frac{\lambda}{4} \phi^4(t,x) \right], \eeq
where potential $V(t,x)$ determines the interaction of the free field and the mirror. In Sec.~\ref{sec:two} we consider the case of two ideal mirrors which corresponds to an infinitely high potential well. In Sec.~\ref{sec:non} we model a nonideal mirror with a delta-functional potential.

We would like to consider a nonhomogeneous  motion of the mirrors, i.e. such potential $V(t,x)$ that changes with time in an arbitrary inertial reference frame. Since the Hamiltonian of such a theory explicitly depends on time, quantum loop corrections to the tree-level correlation functions should be calculated with the nonstationary Schwinger--Keldysh diagrammatic technique~\cite{Schwinger, Keldysh}. In this subsection we briefly introduce this technique. A comprehensive review can be found in~\cite{Akhmedov:dS, Kamenev, Berges, Rammer, Calzetta, Landau:vol10}. 

First of all, suppose we know the state of the system~\eqref{eq:S} at a moment $t_0$ and would like to calculate the expectation value of the operator $\mO$ at a moment $t$. In the interaction picture we have the following expectation value\footnote{If the interaction is turned on and switched off adiabatically and $| in \rangle$ is the true vacuum state of the free theory, then the interaction cannot disturb the $| in \rangle$ state, i.e. $| in \rangle$ and $| out \rangle$ states coincide. In this case in-in expectation value~\eqref{eq:in-in} can be reduced to the in-out expectation value, and Schwinger--Keldysh technique reproduces the standard Feynman technique (see~\cite{Akhmedov:dS} for the details).}:
\beq \label{eq:in-in}
\langle \mO \rangle(t) = \left\langle U^\dagger(\infty, t_0) \mT \left[ \mO(t) U(\infty, t_0) \right] \right\rangle. \eeq
Here $\mT$ denotes the time-ordering and $\langle \cdots \rangle \equiv \langle in | \cdots | in \rangle$, where $| in \rangle$ is the state of the system at the moment $t_0$. The evolution operator in our case has the following form:
\beq U(t_1, t_2) = \mT \exp \left[ -i \frac{\lambda}{4} \int_{t_2}^{t_1} dt \int dx \, \phi^4(t,x) \right]. \eeq

Now it is straightforward to see that loop corrections to an arbitrary tree-level correlation function (i.e. correlation function~\eqref{eq:in-in} with $\mO = \phi(t_1, x_1) \phi(t_2, x_2) \cdots \phi(t_n, x_n)$) reduces to the sum over all possible products of the following four bare propagators:
\beq \label{eq:pm} \begin{gathered}
i G_{12}^{--} \equiv \big\langle \mT \phi_1 \phi_2 \big\rangle = \theta(t_1 - t_2) \left \langle \phi_1 \phi_2 \right\rangle + \theta(t_2 - t_1) \left\langle \phi_2 \phi_1 \right\rangle; \\
i G_{12}^{++} \equiv \big\langle \widetilde{\mT} \phi_1 \phi_2 \big\rangle = \theta(t_2 - t_1) \left\langle \phi_1 \phi_2 \right\rangle + \theta(t_1 - t_2) \left\langle \phi_2 \phi_1 \right\rangle; \\
i G_{12}^{+-} \equiv \left\langle \phi_1 \phi_2 \right\rangle; \qquad i G_{12}^{-+} \equiv \left\langle \phi_2 \phi_1 \right\rangle.
\end{gathered} \eeq
Here $\widetilde{\mT}$ is the reverse time-ordering and we denote $G_{12} = G(t_1, x_1; t_2, x_2)$, $\phi_i = \phi(t_i, x_i)$ for shortness. In this notation ``$+$" and ``$-$" signs mean that $\phi$-operators come from $U^\dagger$ (i.e. from the anti-time-ordered part of the full propagator) or from $U$ (i.e. from the time-ordered part), respectively. Similarly, ``$+$" and ``$-$" vertices come from $U^\dagger$ and $U$ and ascribe to the diagram factors $i \frac{\lambda}{4}$ and $-i \frac{\lambda}{4}$, respectively. For example, the one-loop correction to the $G_{12}^{+-}$ propagator is as follows:
\beq \Delta G_{12}^{+-} = 3 i \lambda \int_{t_0}^\infty dt_3 \, G_{13}^{++} \left( G_{33}^{++} \right)^2 G_{32}^{+-} - 3 i \lambda \int_{t_0}^\infty dt_3 \, G_{13}^{+-} \left( G_{33}^{--} \right)^2 G_{32}^{--}. \eeq
Note that we excluded disconnected diagrams because in Schwinger--Keldysh technique such diagrams always cancel each other. Also we took into account symmetry factor (in this case it is 12) and included the imaginary unit $i$ into the definition of propagators.

It is easy to see that propagators~\eqref{eq:pm} obey the relation $G^{++} + G^{--} = G^{+-} + G^{-+}$. Hence, it is convenient to perform the Keldysh rotation~\cite{Keldysh, Kamenev, Berges} and introduce the Keldysh, retarded and advanced propagators:
\beq \label{eq:KRA} \begin{aligned}
G_{12}^K &= \frac{1}{2} \left( G_{12}^{+-} + G_{12}^{-+} \right) = \frac{1}{2} \left\langle \left\{ \phi_1, \phi_2 \right\} \right\rangle; \\
G_{12}^{R/A} &= G_{12}^{--} - G_{12}^{\mp \pm} = \pm \theta(\pm t_1 \mp t_2) \left\langle \left[ \phi_1, \phi_2 \right] \right\rangle.
\end{aligned} \eeq
These propagators have a simple physical interpretation. On the one hand, retarded and advanced propagators describe the propagation of some localized perturbations (e.g. particle or quasi-particle). Hence, at the tree level they do not depend on the state of the system (note that this agrees with the definition~\eqref{eq:KRA}, because commutator is a c-number). On the other hand, Keldysh propagator contains the information about the state of the system:
\beq \label{eq:K}
G_{12}^K = \int \frac{dp}{2\pi} \int \frac{dq}{2\pi} \left[ \left( \frac{1}{2} \delta_{pq} + \langle a_p^\dagger a_q \rangle \right) g_p^*(t_1, x_1) g_q(t_2, x_2) + \langle a_p a_q \rangle g_p(t_1, x_1) g_q(t_2, x_2) + h.c. \right], \eeq
where $\delta_{pq} \equiv \delta(p - q)$, $a_p^\dagger$ and $a_p$ are bosonic creation and annihilation operators, and $g_p(t,x)$ is the mode function in the free field decomposition $\phi(t,x) = \int \frac{dp}{2\pi} \left[ a_p g_p(t,x) + a_p^\dagger g_p^*(t,x) \right]$. Usually expectation values $n_{pq} \equiv \langle a_p^\dagger a_q \rangle$ and $\kappa_{pq} \equiv \langle a_p a_q \rangle$ are diagonal in momenta, $n_{pq} = n_p \delta_{pq}$, $\kappa_{pq} = \kappa_p \delta_{p,-q}$. In this case quantities $n_p$ and $\kappa_p$ are called energy level density and anomalous quantum average, respectively. In the case when mode functions $g_p$ are not pure exponentials, these definitions should also take into account the change in the modes (see the discussion in the Sec.~\ref{sec:discussion}). However, we will apply the same terminology to the quantities $n_{pq}$ and $\kappa_{pq}$ as well.

Note that for a vacuum in-state, $a_p | in \rangle = 0$, both $n_{pq} = 0$ and $\kappa_{pq} = 0$. At the tree-level they remain zero during time evolution. Therefore, particle production is related only to the change in the \textit{modes}. However, in nonstationary situation these vacuum expectation values can also receive nonzero loop corrections. This would indicate the change in the \textit{state} and refute the semiclassical approximation.

The most interesting case of nonzero loop corrections is the case of so-called secular growth when $n_{pq}$ and/or $\kappa_{pq}$ indefinitely grow in the limit $T = \frac{t_1 + t_2}{2} \rightarrow \infty$, $\Delta t = t_1 - t_2 = \const$ ($t_1$ and $t_2$ are external times of the exact Keldysh propagator). Such a growth means that at some moment ($T \sim 1/\lambda$) loop corrections exceed tree-level values even for an infinitesimal coupling constant. After this moment the perturbation theory becomes inapplicable, so some kind of resummation must be performed in order to estimate the exact correlation functions. Such a resummation also allows to find the correct final state of the system after interactions are turned off~\cite{Akhmedov, Burda, Polyakov, Akhmedov:dS, Popov, Pavlenko, Akhmedov:Et, Akhmedov:Ex}.

Besides that, the Keldysh propagator is closely connected with the stress-energy flux. In a gaussian theory the relation is:
\beq \label{eq:T}
\langle T_{tx} \rangle = \pd_{t_1} \pd_{x_2} G_{12}^K \big|_{t_1 = t_2, x_1 = x_2}. \eeq
On the tree level nonzero flux appears due to the amplification of zero-mode fluctuations, i.e. due to the fact that mode functions $g_p(t,x)$ do not coincide with a simple exponential function. This effect was observed and discussed in many seminal papers, e.g.~\cite{Birrell, Davies:1976, Davies:1977, DeWitt}. However, we would like to emphasize that quantum loop corrections also can make a significant contribution to the stress-energy flux. Indeed, the unlimited time growth of $n_{pq}$ and $\kappa_{pq}$ in the decomposition~\eqref{eq:K} inevitably generates a valid contribution to the stress-energy flux~\eqref{eq:T}. Moreover, for large times loop corrections start to dominate. This is the other reason to study the secular growth of quantum averages.

Due to these reasons in this paper we calculate loop corrections in the limit $T \gg \Delta t$. In addition, we consider small coupling constants, $\lambda \rightarrow 0$, $\lambda T = \const$, in order to single out the leading quantum corrections to the tree-level propagators. This limit has two important implications. First, loop corrections to the retarded and advanced propagators are negligible, because they do not grow in the limit $T \rightarrow \infty$. For example, the first loop correction to the retarded propagator is as follows:
\beq \Delta G_{12}^R = -3 i \lambda \int_{t_0}^\infty dt_3 G_{13}^R G_{33}^K G_{32}^R = -3 i \lambda \int_{t_2}^{t_1} dt_3 G_{13}^R G_{33}^K G_{32}^R \sim \mO\left( \lambda T^0 \right). \eeq
For the second identity we used the causal structure of the retarded propagator, i.e. took into account theta-functions in the definition~\eqref{eq:KRA}. Obviously, due to the specific limits of integration this expression cannot grow in the limit $T \gg \Delta t$ for a fixed $\Delta t$. The growth as $\Delta t \rightarrow \infty$ may affect properties of quasi-particles, but it does not affect the state of the entire system or the stress-energy flux~\cite{Pavlenko}. Moreover, it is easy to check that loop corrections do not change the causal structure of retarded and advanced propagators. Therefore, higher-loop corrections possess the same behavior~\cite{Akhmedov:dS, Kamenev, Berges, Rammer, Calzetta, Landau:vol10}. Hence, we can neglect them in the limit in question.

Second, in the limit in question leading loop corrections to the energy level density and anomalous quantum average in the decomposition~\eqref{eq:K} depend only on the average time:
\beq \label{eq:evolved} \begin{aligned}
n_{pq} &= \left\langle U^\dagger(\infty, t_0) \mT \left[ a_p^\dagger(t_1) a_q(t_2) U(\infty, t_0) \right] \right\rangle \approx \left\langle U^\dagger(\infty, t_0) \mT \left[ a_p^\dagger(T) a_q(T) U(\infty, t_0) \right] \right\rangle + \cdots, \\
\kappa_{pq} &= \left\langle U^\dagger(\infty, t_0) \mT \left[ a_p(t_1) a_q(t_2) U(\infty, t_0) \right] \right\rangle \approx \left\langle U^\dagger(\infty, t_0) \mT \left[ a_p(T) a_q(T) U(\infty, t_0) \right] \right\rangle + \cdots,
\end{aligned} \eeq
where ellipsis denote the subleading contributions in the limit $\lambda \rightarrow 0$, $T \gg \Delta t$, $T \sim 1/\lambda$. Hence, expressions $n_{pq}(T)$ and $\kappa_{pq}(T)$ can be interpreted as the exact energy level density and anomalous quantum average at the moment $T$.

Thus, in this paper we calculate loop corrections to the tree-level quantum averages in the theory~\eqref{eq:S} and show that they possess secular growth. We start with the two-loop (``sunset") corrections, which are the simplest nontrivial loop corrections. It is straightforward to show that in the limit $\lambda \rightarrow 0$, $T \gg \Delta t$, $T \sim 1/\lambda$ these corrections are given by the following formulae:
\begin{align}
\label{eq:n} n_{pq}(T) &\approx 2 \lambda^2 \int d^2 x_1 d^2 x_2 \, \theta\left(T - t_1\right) \theta\left(T - t_2\right) g_{p,1} g_{q,2}^* \left[ \int \frac{d k}{2 \pi} g_{k,1} g_{k,2}^* \right]^3, \\
\label{eq:k} \kappa_{pq}(T) &\approx -2 \lambda^2 \int d^2 x_1 d^2 x_2 \, \theta\left(T - t_1\right) \theta\left(t_1 - t_2\right) \left[ g_{p,1}^* g_{q,2}^* + g_{p,2}^* g_{q,1}^* \right] \left[ \int \frac{d k}{2 \pi} g_{k,1} g_{k,2}^* \right]^3,
\end{align}
where we denoted $g_{p,i} = g_p(t_i, x_i)$ for shortness. In the following sections we determine the exact modes for several types of the potential $V(t,x)$ in the theory~\eqref{eq:S}, substitute them into the identities~\eqref{eq:n} and~\eqref{eq:k} and show that both $n_{pq}(T) \sim (\lambda T)^2$ and $\kappa_{pq}(T) \sim (\lambda T)^2$. We emphasize that these functions should be considered as parts of the Keldysh propagator~\eqref{eq:K} which has a more fundamental meaning.

\subsection{Physical picture}
\label{sec:picture}

It is useful to keep in mind the following physical picture related to the scalar DCE. Consider electromagnetic field interacting with a thin layer of cold electronic plasma. Fixing the Lorentz gauge and ignoring transverse effects\footnote{I.e. neglecting interactions between transverse and longitudinal modes and setting transverse momentum $k_\bot = 0$. The opposite case reproduces a massive scalar field with mass $m^2 = k_\bot^2$.} we obtain the following equation on the transverse component of the vector potential $A_z(t,x)$ \cite{Neto, Bulanov-1}:
\beq \frac{\pd^2 A_z(t,x)}{\pd t^2} - c^2 \frac{\pd^2 A_z(t,x)}{\pd x^2} + \frac{\omega_{pe}^2(t,x)}{\gamma(t,x)} A_z(t,x) = 0, \qquad \omega_{pe}^2(t,x) = \frac{4 \pi e^2 n(t,x)}{m_e}. \eeq
Here $c$ is the speed of light, $\omega_{pe}$ is Langmuir frequency, $e$ and $m_e$ are electron charge and mass, $n(t,x)$ is electron density distribution and $\gamma(t,x)$ is the Lorentz factor of the plasma sheet. We assume that plasma sheet moves along the X-axis, so that function $x = x(t)$ describes its position at the moment $t$. Approximating the electron density distribution by Dirac delta-function, $n(t,x) = n_0 l \delta\left[x - x(t)\right]$, where $n_0$ is the average electron density and $l$ is the thickness of the electron layer, we obtain the equation on a two-dimensional massless scalar field:
\beq \label{eq:non-ideal}
\left[ \frac{\pd^2}{\pd t^2} - c^2 \frac{\pd^2}{\pd x^2} - \frac{\alpha}{\gamma(t,x)} \delta\left[ x - x(t) \right] \right] \phi(t,x) = 0, \eeq
where $\phi(t,x) = A_z(t,x) \sqrt{S_\bot}$, $\alpha = 4 \pi e^2 n_0 l/m_e$ and $S_\bot$ is the area of mirrors (this factor does not affect equations of motion, but it is necessary for the correct dimensional reduction). As a rough approximation we can set $l \sim c / \omega_{pe}$, which yields $\alpha \sim c \omega_{pe}$. A typical metal mirror has $\omega_{pe} \sim 10^{16}$ s$^{-1}$, i.e. $\alpha/c^2 \sim 10^5$ cm$^{-1}$. Practically such mirror also can be implemented by a breaking plasma wakewave with typical parameters $n_0 \sim 10^{17}$ cm$^{-3}$ and $l \sim 10^{-2}$ cm; in this case $\alpha / c^2 \sim 10^4$ cm$^{-1}$ \cite{Bulanov-1, Bulanov-2}. Yet another way to implement rapidly moving mirror involves superconducting circuits which approximately reproduce the model~\eqref{eq:non-ideal} with $\alpha / c^2 \sim 10$ cm$^{-1}$ \cite{Johansson:2009, Johansson:2010, Nation, Wilson}.

Note that in the limit $\alpha \rightarrow \infty$ eq.~\eqref{eq:non-ideal} reproduces the case of the ideal mirror:
\beq \label{eq:ideal}
\left[ \frac{\pd^2}{\pd t^2} - c^2 \frac{\pd^2}{\pd x^2} \right] \phi(t,x) = 0, \quad \phi\left[t, x(t)\right] = 0. \eeq
Moreover, the transmission coefficient of a nonideal mirror is proportional to $\omega / \alpha$ if $\omega \ll \alpha$, where $\omega$ is the energy of an incident wave. A derivation of this statement can be found in Sec.~\ref{sec:non-tree}. Therefore, at such energy scales we can use a simplified eq.~\eqref{eq:ideal} instead of more accurate eq.~\eqref{eq:non-ideal}.

Finally, recall that we would like to consider quantum corrections to free scalar theories~\eqref{eq:non-ideal} and~\eqref{eq:ideal}. We choose $\lambda \phi^4$ as the simplest example of a non-Gaussian theory (see Sec.~\ref{sec:SK}). On the one hand, such interaction can be interpreted as a toy model of low-energy effective QED~\cite{Heisenberg,Schwinger:1951}, although there is no direct correspondence between these theories. On the other hand, in the context of circuit QED $\lambda \phi^4$ theory describes a nonlinear waveguide~\cite{Johansson:2010, Yurke}. In particular, such a nonlinear self-interacting term can be simulated by embedding a SQUID in the transmission line~\cite{Nation:2008, Bourassa, Boutin, Buks, Blais}. In this case the dimensionless coefficient of nonlinearity (the ratio of the Kerr coefficient and characteristic energy) can be estimated as $K \sim 10^{-6}$. In our terminology this coefficient corresponds to $K \sim \lambda / \omega_0^2 \sim \lambda \Lambda^2$ where $\omega_0$ is the characteristic energy of oscillations and $\Lambda$ is the characteristic size of the system. We expect that such nonlinear effects can be combined with the reflecting boundary conditions and implemented using superconducting circuits similarly to the tree-level measurements~\cite{Wilson, Lahteenmaki}.

In the remainder of this paper we use units $\hbar = c = 1$ unless otherwise specified. Also we assume the $(+,-)$ signature for the metric tensor.

\section{Two ideal mirrors}
\label{sec:two}

First of all, note that the theory of a massless scalar field with a single mirror considered in~\cite{Alexeev} does not have a natural IR cut-off. This makes it difficult to distinguish between the true secular growth and standard IR divergences of loop integrals. For example, such artifacts arise in a pure stationary theory at small evolution times (see App.~\ref{sec:artifacts}). In this section we study this issue by adding the second mirror at a finite distance\footnote{Another possibility to impose a natural IR cut-off is to consider fields with nonzero physical mass. This approach was investigated in~\cite{Astrakhantsev}. However, the equations of motion for a massive scalar field on a nonhomogeneously moving mirror background are very difficult to solve. Due to the same reason loop corrections to $n_{pq}$ and $\kappa_{pq}$ are very difficult to calculate analytically in this case.}. For simplicity we assume that both mirrors are perfectly reflecting.

We confirm that at small evolution times quantum averages in the two-mirror problem exhibit an unphysical quadratic growth which can be eliminated by the proper regularization (Sec.~\ref{sec:sim-small}). However, at large evolution times this unphysical growth is replaced by a physically meaningful secular growth which indicates a change in the state of the theory. On the one hand, this growth is associated with the violation of conformal invariance by the $\lambda \phi^4$ interaction term. On the other hand, it is closely related to the violation of the energy-conservation law on a nonstationary background (i.e. to the pumping of energy by an external force). We also reproduce the results of~\cite{Alexeev} in the limit of infinitely distant mirrors.

In this section we mainly work with ``broken'' mirror trajectories. Physically, such a trajectory corresponds to a single sudden kick which results in a discontinuous change in the velocity of the mirror. We assume that both mirrors move along ``broken'' trajectories. Moreover, we would like to consider asymptotically stationary motions, so we need to equate the final velocities of the mirrors. Note that we also need to adjust the moments when the kicks are applied to the mirrors. In Secs.~\ref{sec:sim} and~\ref{sec:synch} we consider two of the most natural adjustment options. In Sec.~\ref{sec:cavity} we also discuss periodically oscillating mirror trajectories which model a resonant cavity.

\subsection{Geometrical method of calculating modes}
\label{sec:geometry}

In this subsection we discuss the quantization of a massless scalar field on the background of two perfectly reflective mirrors:
\beq \label{eq:motion}
\pd_\mu \pd^\mu \phi = 0, \quad \phi\left[t, L(t)\right] = \phi\left[t, R(t)\right] = 0, \eeq
where $L(t)$ and $R(t)$ denote the position of the left and right mirror at the moment $t$, respectively. We assume that mirrors are at rest before the moment $t = 0$, i.e. $L(t < 0) = 0$ and $R(t < 0) = \Lambda$. As was shown in~\cite{Moore, Davies:1976, Birrell}, the quantized field is represented by the following mode decomposition:
\beq \phi(t,x) = \sum_{n=1}^\infty \left[ a_n g_n(t,x) + a_n^\dagger g_n^*(t,x) \right], \eeq
where $\left[ a_m, a_n^\dagger \right] = \delta_{mn}$. In the stationary case ($L(t) = 0$, $R(t) = \Lambda$ for all $t$) the $n=1$ mode corresponds to a standing wave with the frequency $\omega_1 = \frac{\pi}{\Lambda}$. This is the lowest energy excitation of the cavity. The mode functions for an arbitrary motion of the mirrors can be written in terms of two auxiliary functions $G(z)$ and $F(z)$:
\beq \label{eq:mode-moore}
g_n(t,x) = \frac{i}{\sqrt{4 \pi n}} \left[ e^{-i \pi n G(t + x)} - e^{-i \pi n F(t - x)} \right], \eeq
which satisfy the generalized Moore's equations:
\beq \label{eq:GM}
G\left[ t + L(t) \right] - F\left[ t - L(t) \right] = 0, \quad G\left[ t + R(t) \right] - F\left[ t - R(t) \right] = 2. \eeq
In a stationary case these functions can be easily found to be $G(z) = F(z) = \frac{z}{\Lambda}$. For arbitrary trajectories $L(t)$ and $R(t)$ equations~\eqref{eq:GM} can be solved recursively by the geometrical method proposed in~\cite{Cole, Meplan} and extended in~\cite{Li}. We briefly discuss the method in this subsection and apply it to several types of motions in Secs.~\ref{sec:sim}, \ref{sec:synch} and~\ref{sec:cavity}. Recall that we assume both mirrors are at rest before the moment $t = 0$. This implies $G(z \le \Lambda) = F(z \le 0) = \frac{z}{\Lambda}$. Therefore, it is convenient to define $G$ static region ($t+x \le \Lambda$) and $F$ static region ($t-x \le 0$) where the corresponding functions are fixed, $G(t+x) = \frac{t + x}{\Lambda}$, $F(t-x) = \frac{t - x}{\Lambda}$.

The general idea of the geometrical method is to trace back functions $G(z)$ and $F(z)$ along the sequence of null lines until a null line intersects the time axis in a static region. In other words, this method uses that functions $G(t+x)$ and $F(t-x)$ from the decomposition~\eqref{eq:mode-moore} are constant on the null lines $t + x = \const$ and $t - x= \const$. This allows to trace functions $G$ and $F$ along a null line between the mirrors. At the same time, eqs.~\eqref{eq:GM} relate functions $G$ and $F$ on reflected null lines. This, in turn, allows us to proceed until a null line reaches a static region where $G$ or $F$ is known.

Let us illustrate this idea on the function $G(z)$ (see Fig.~\ref{fig:two-mirrors}). First, we draw a null line from the point $(z,0)$ until it intersects the right mirror at the point $\left( t_1, R(t_1) \right)$:
\beq z = t_1 + R(t_1), \quad \text{hence}, \quad G(z) = G\left( t_1 + R(t_1) \right). \eeq
Then we relate functions $G$ and $F$ on the right mirror:
\beq G(z) = G\left( t_1 + R(t_1) \right) = F\left( t_1 - R(t_1) \right) + 2, \eeq
and draw a null line from the point $\left( t_1, R(t_1) \right)$ to $\left( t_2, L(t_2) \right)$:
\beq t_1 - R(t_1) = t_2 - L(t_2), \quad \text{hence}, \quad F\left( t_1 - R(t_1) \right) = F\left( t_2 - L(t_2) \right). \eeq
Finally, we switch back to the function $G$ and find the next intersection of the null line and the right mirror:
\beq t_2 + L(t_2) = t_3 + R(t_3), \quad \text{hence}, \quad F\left( t_2 - L(t_2) \right) = G\left( t_2 + L(t_2) \right) = G\left( t_3 + R(t_3) \right). \eeq
This defines the step of the recursion:
\beq G(z) = G\left( t_1 + R(t_1) \right) = G\left( t_3 + R(t_3) \right) + 2 = G\left( t_5 + R(t_5) \right) + 4 = \cdots. \eeq
Note that the value of $G(z)$ increases by 2 every time the null line hits the right mirror. This process is terminated only when a null line enters a static region where $F(z)$ or $G(z)$ can be evaluated explicitly. There are two possible ways to enter such a region. First, a null line reflecting off the right mirror can enter the $F$ static region. Second, a null line reflecting off the left mirror can enter the $G$ static region. In both cases function $G(z)$ reduces to the following expression:
\beq \label{eq:G}
G(z) = 2 n + \frac{t_\text{final}}{\Lambda}, \eeq
where $n$ is the number of reflections off the right mirror and $t_\text{final}$ is the moment at which the last null line intersect the time-axis in the static region (see Fig.~\ref{fig:two-mirrors}):
\begin{figure}[t]
\center{\includegraphics[scale=0.4]{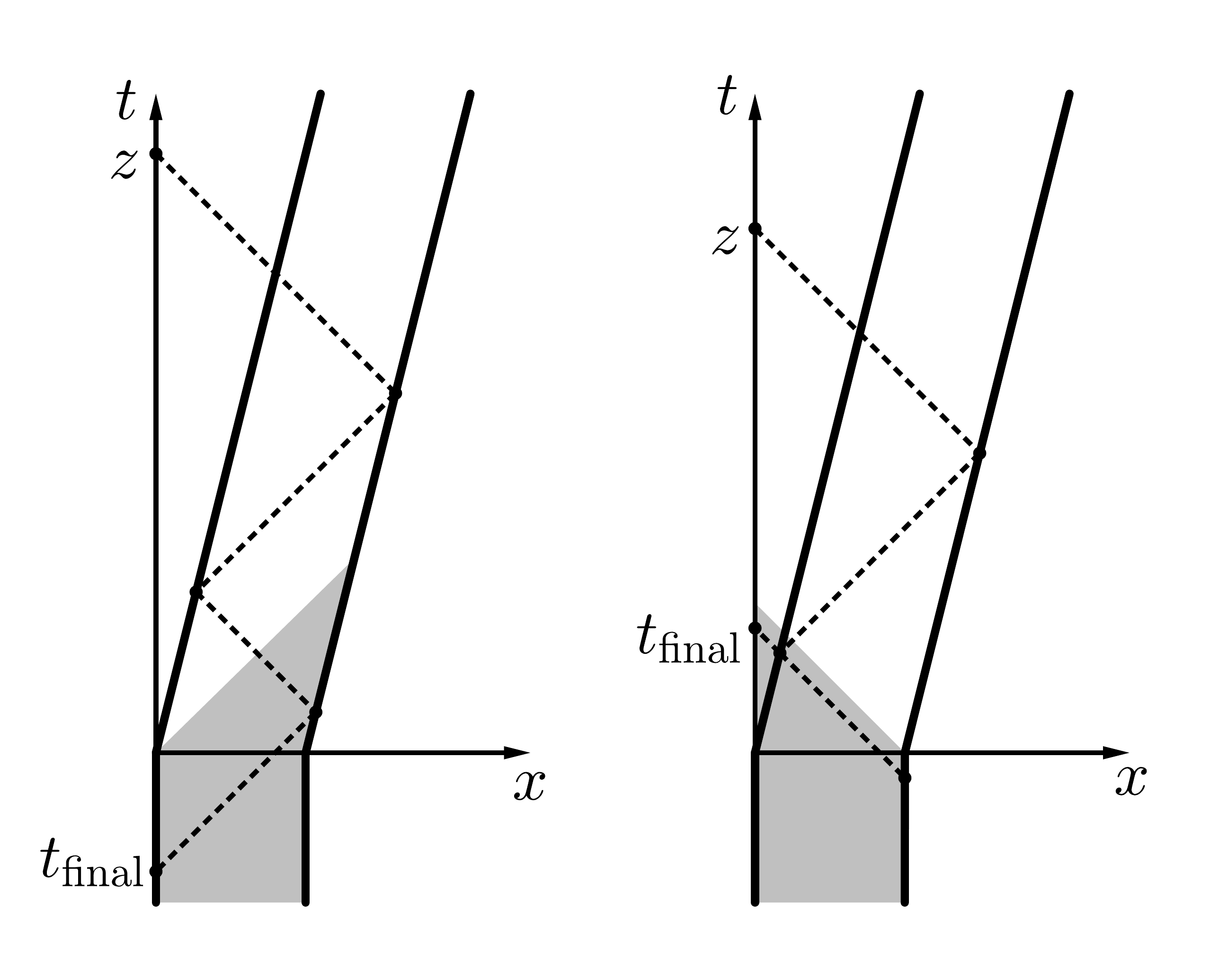}}\caption{Calculation of the $t_\text{final}$ for the $F$ static region (to the left) and $G$ static region (to the right). Bold solid lines denote word-lines of the mirrors, dashed lines denote null lines, and light gray areas denote $F$ or $G$ static regions.}\label{fig:two-mirrors}
\end{figure} 
\beq t_\text{final} = \begin{cases} z - 2 \left[ \sum_{i=1}^n R\left( t_{2 i-1} \right) - \sum_{i=1}^{n-1} L\left( t_{2 i} \right) \right], \quad & \text{for $F$ static region}, \\
z - 2 \left[ \sum_{i=1}^n R\left( t_{2 i-1} \right) - \sum_{i=1}^n L\left( t_{2 i} \right) \right], \quad & \text{for $G$ static region}.
\end{cases} \eeq
Note that for time-like mirror world-lines moments $t_1, t_2, \cdots, t_\text{final}$ always exist and are unique~\cite{Davies:1976, Cole}. Hence, the identity~\eqref{eq:G} correctly restores the function $G(z)$ for all values of $z$. Also note that function~\eqref{eq:G} is continuous because function $t_\text{final}(z)$ decreases by $2 \Lambda$ at points where the number of reflections $n(z)$ increases by one.

The function $F(z)$ can be restored using the similar procedure~\cite{Li}. However, for relatively simple functions $L(t)$ it may be more convenient to use the identity $F\left[ t - L(t) \right] = G\left[ t + L(t) \right]$. In this case it is sufficient to find the inverse function of $z(t) = t - L(t)$ (provided that $G(z)$ is known).

Finally, let us discuss the possible choices for the functions $L(t)$ and $R(t)$. We would like to consider asymptotically uniform trajectories, i.e. trajectories with fixed velocities $\dot{L}(\pm \infty) = \dot{R}(\pm \infty) = \beta_\pm$, $|\beta_\pm| < 1$. Such trajectories describe a cavity that is approximately stationary in the infinite past and future but undergoes expansion or contraction at intermediate times. For convenience we choose a coordinate system where both mirrors are at rest ($\beta_- = 0$) before the moment $t = 0$. 

The simplest example of such a nonstationary motion is a combination of so-called ``broken'' trajectories each of which describes a single discontinuous change in the velocity of the mirror:
\beq \label{eq:broken}
x(t) = x(t_x) + \beta (t - t_x) \theta(t - t_x), \eeq
where $x = L,R$ and $t_x$ is the moment when the corresponding mirror experiences a sudden kick. We remind that $L(t_L) = 0$ and $R(t_R) = \Lambda$. The case $t_L = t_R = 0$ and $\beta = \frac{1}{4}$ is depicted on the Fig.~\ref{fig:two-mirrors}. This trajectory can be considered as an approximation to a finite-period motion with constant proper acceleration $w$:
\beq \label{trj1} x(t) - x(t_x) = \begin{cases} \frac{1}{w}\left( \sqrt{1 + w^2 (t - t_x)^2} - 1 \right), \quad  &\text{for} \quad 0 < t - t_x < \frac{\gamma \beta}{w}, \\ \frac{1}{w}\left( \frac{1}{\sqrt{1 - \beta^2}} - 1 \right) + \beta (t - t_x), \quad &\text{for} \quad t - t_x > \frac{\gamma \beta}{w}, \end{cases} \eeq
or to an eternally accelerated motion with velocity exponentially close to $\beta$:
\beq \label{trj2} x(t) - x(t_x) = \beta (t - t_x) - \frac{\beta}{w} \left( 1 - e^{-w (t - t_x)} \right), \quad \text{for} \quad t > t_x. \eeq
Both of these trajectories reproduce~\eqref{eq:broken} in the limit $w \rightarrow \infty$ and smoothly connect asymptotically uniform regions. For the tree-level calculations this difference may be crucial, because discontinuity in the velocity generates a singular stress-energy flux (e.g. see~\cite{Davies:1976, Davies:1977, Castagnino}). At the same time, in the following subsection we will argue that loop-level calculations coincide for all asymptotically uniform mirror trajectories (except the case $|\beta| = 1$ which cannot be considered in our approach). As a result, two-loop corrections to $n_{pq}$ and $\kappa_{pq}$ quadratically grow with time for all such trajectories, although the prefactors of this growth depend on the intermediate motion. Hence, we can use a simple ``broken'' trajectory~\eqref{eq:broken} to illustrate the key points of the calculation.

\subsection{Secular growth as a consequence of the conformal invariance violation}
\label{sec:conformal}

The solution~\eqref{eq:mode-moore} was inspired by the conformal invariance of the free massless scalar field~\cite{Moore, Davies:1976}. In fact, it is easy to check that the following conformal transformation:
\beq \label{eq:conformal}
w + s = G(t + x), \quad w - s = F(t - x), \eeq
reduces the problem~\eqref{eq:motion} to the stationary problem with both mirrors at rest:
\beq \pd_\mu \pd^\mu \phi = 0, \quad \phi(w, s = 0) = \phi(w, s = 1) = 0, \eeq
which immediately implies the solution~\eqref{eq:mode-moore}. However, the $\lambda \phi^4$ interaction breaks the conformal invariance down. In this subsection we will show that this breakdown manifests itself in the secular growth of loop corrections. For simplicity we will assume that the velocities of the mirrors coincide at large times, i.e. $\dot{L}(t) = \dot{R}(t) = \beta$ after some time $t_*$. The reasons for this (physically meaningful) requirement will be explained below.

First, let us remind that two-loop corrections to the energy level density and anomalous quantum average in the $\lambda \phi^4$ theory are given by the following expressions:
\begin{align}
\label{eq:n-two} n_{pq}(T) &\approx 2 \lambda^2 \int_{t_0}^T dt_1 \int_{L(t_1)}^{R(t_1)} dx_1 \int_{t_0}^T dt_2 \int_{L(t_2)}^{R(t_2)} dx_2 \sum_{m,n,k = 1}^\infty I_{p,m,n,k}(t_1, x_1) I_{q,m,n,k}^*(t_2, x_2), \\ 
\label{eq:k-two} \kappa_{pq}(T) &\approx -2 \lambda^2 \int_{t_0}^T dt_1 \int_{L(t_1)}^{R(t_1)} dx_1 \int_{t_0}^{t_1} dt_2 \int_{L(t_2)}^{R(t_2)} dx_2 \sum_{m,n,k = 1}^\infty \left( I_{p,m,n,k}^c(t_1, x_1) I_{q,m,n,k}^*(t_2, x_2) + (p \leftrightarrow q) \right),
\end{align}
where we introduced for shortness
\beq \begin{aligned}
I_{p,m,n,k}(t_1, x_1) &= g_p(t_1, x_1) g_m(t_1, x_1) g_n(t_1, x_1) g_k(t_1, x_1), \\
I_{p,m,n,k}^c(t_1, x_1) &= g_p^*(t_1, x_1) g_m(t_1, x_1) g_n(t_1, x_1) g_k(t_1, x_1).
\end{aligned} \eeq
We would like to estimate these integrals using the conformal transformation~\eqref{eq:conformal}. As we explained in the Sec.~\ref{sec:SK}, we keep only the leading contributions in the limit $\lambda \rightarrow 0$, $T \rightarrow \infty$. Integrands of~\eqref{eq:n-two} and~\eqref{eq:k-two} consist of oscillating exponents, so we can estimate $| I_{p,m,n,k}(t,x) | < \frac{1}{\pi^4}$ and $| I^c_{p,m,n,k}(t,x) | < \frac{1}{\pi^4}$. This implies that the integration over the regions with finite areas cannot provide growing contributions to~\eqref{eq:n-two} and~\eqref{eq:k-two}. Such contributions are negligible in the limit in question; therefore, such integrations can be excluded. Due to the same reason we can set $t_0 = 0$ (the integrals cannot indefinitely grow in the limit $t_0 \rightarrow -\infty$ because mirrors are stationary in the past, see App.~\ref{sec:artifacts}). This rationale allows us to simplify the integrals and single out the leading, growing with time contributions to~\eqref{eq:n-two} and~\eqref{eq:k-two}. This is the only type of contributions which survive in the limit in question.

After the conformal transformation~\eqref{eq:conformal} we obtain the following integrals:
\beq \int_0^T dt_1 \int_{L(t_1)}^{R(t_1)} dx_1 \, I_{p,m,n,k}(t_1, x_1) \approx \int_0^{G\left[ T + L(T) \right]} dw \int_0^1 ds \, g'(w+s) f'(w-s) I_{p,m,n,k}(w,s), \eeq
where $g'(w+s) f'(w-s) = \frac{d G^{-1}(z)}{dz} \Big|_{z = w + s} \frac{d F^{-1}(z)}{dz} \Big|_{z = w - s}$ is the conformal factor, $I_{p,m,n,k}(w,s) = h_p(w,s) h_m(w,s) h_n(w,s) h_k(w,s)$ and $h_p(w,s) = \frac{i}{\sqrt{4 \pi p}} \left[ e^{-i \pi p (w + s)} - e^{-i \pi p (w - s)} \right]$. The integrals of $I^*_{p,m,n,k}(t,x)$ and $I^c_{p,m,n,k}(t,x)$ have the same structure. Note that $g'(w+s)$ and $f'(w-s)$ are positive if we consider space-like mirror word-lines with $| \dot{L}(t)| < 1$ and $| \dot{R}(t)| < 1$.

Finally, let us make yet another change and introduce the coordinates $u = w - s$, $v = w + s$:
\beq \label{eq:int-uv}
\int_0^T dt_1 \int_{L(t_1)}^{R(t_1)} dx_2 \, I_{p,m,n,k}(t_1, x_1) \approx \int_0^{G\left[T + L(T) \right]} du \int_u^{u+2} dv \, \frac{1}{2} g'(v) f'(u) I_{p,m,n,k}(u,v). \eeq
For general trajectories $L(t)$ and $R(t)$ this integral is very complex, but in some physically meaningful cases it is significantly simpler that the initial integrals~\eqref{eq:n-two} and~\eqref{eq:k-two}.

Namely, assume that the velocities of the mirrors coincide for large times, i.e. $\dot{L}(t) = \dot{R}(t) = \beta$ for $t > t_*$. In this case the geometrical picture of Sec.~\ref{sec:geometry} implies that functions $G(z)$ and $F(z)$ periodically grow, i.e. $G(z + \Delta z_G) = G(z) + 2$, $F(z + \Delta z_F) = F(z) + 2$. In the geometrical language of Sec.~\ref{sec:geometry} the increase of the argument by $\Delta z_G = \frac{2 \Lambda_*}{1 - |\beta|}$ or $\Delta z_F = \frac{2 \Lambda_*}{1 + |\beta|}$ adds an additional light ray reflection cycle to the derivation of $G$ or $F$, respectively; $\Lambda_*$ is the distance between the mirrors at the reference frame for $t > t_*$. Hence, starting from some moment $y_*$ the inverse functions $g(y)$ and $f(y)$ also periodically grow, i.e. $g(y + 2) = g(y) + \Delta z_G$, $f(y + 2) = f(y) + \Delta z_F$. What is even more important, their derivatives are simply periodic: $g'(y + 2) = g'(y)$, $f'(y + 2) = f'(y)$. Therefore, they can be expanded into a Fourier series:
\beq \label{eq:Fourier} \begin{gathered}
g'(y) = g'\left[ y - y_* - 2 n(y) \right] = \sum_{n = -\infty}^\infty g_n e^{i \pi n y}, \quad g_n = \frac{1}{2} \int_0^2 g'(y) e^{- i \pi n y} dy, \\
f'(y) = f'\left[ y - y_* - 2 n(y) \right] = \sum_{n = -\infty}^\infty f_n e^{i \pi n y}, \quad f_n = \frac{1}{2} \int_0^2 f'(y) e^{- i \pi n y} dy,
\end{gathered} \eeq
where $n(y) \in \mathbb{N}$ and $0 < y - y_* - n(y) < 2$. This periodicity implies that the integral over $dv$ in~\eqref{eq:int-uv} has contributions that do not depend on $u$:
\beq \int_u^{u+2} dv g'(v) e^{-i \pi (p + m + n + k) v} = \int_u^{u+2} dv \sum_{l = -\infty}^\infty g_n e^{i \pi l v} e^{-i \pi (p+m+n+k) v} = 2 g_{p + m + n + k}. \eeq
Therefore, the integral~\eqref{eq:int-uv} indefinitely grows with time, provided that $g_{p + m + n + k} \neq 0$ and $T > t_*$:
\beq \int_{G\left[ t_* + L(t_*) \right]}^{G\left[T + L(T) \right]} du f'(u) g_{p + m + n + k} \sim g_{p + m + n + k} \left[ T - L(T) - t_* + L(t_*) \right]. \eeq
This, in turn, immediately implies the secular growth of $n_{pq}$ and $\kappa_{pq}$:
\begin{align}
\label{eq:n-two-fin} n_{pq}(T) &\approx 2 \lambda^2 \frac{\left[ T - L(T) - t_* + L(t_*) \right]^2}{(4 \pi)^4 \sqrt{p q}} \sum_{m,n,k=1}^\infty \frac{g_{p + m + n + k} \, g_{q + m + n + k}^*}{m n k}, \\
\label{eq:k-two-fin} \kappa_{pq}(T) &\approx -\lambda^2 \frac{\left[ T - L(T) - t_* + L(t_*) \right]^2}{(4 \pi)^4 \sqrt{p q}} \sum_{m,n,k=1}^\infty \frac{g_{-p + m + n + k} \, g_{q + m + n + k}^* + (p \leftrightarrow q)}{m n k}.
\end{align}
We remind that we keep only the leading expressions in the limit $\lambda \rightarrow 0$, $T \gg \Delta t$, $T \sim 1/\lambda$. The oscillating contributions and contributions of the form $\lambda^2 T^\alpha$ with $\alpha < 2$ simply die out in this limit. Note that the sums over the virtual momenta are convergent because $g_n \sim \frac{\Lambda}{n}$ for $n \gg 1$:
\beq \sum_{m,n,k=1}^\infty \frac{g_{p + m + n + k} g_{q + m + n + k}}{mnk} \sim \sum_{m,n,k=1}^\infty \frac{\Lambda^2}{(p+m+n+k) (q+m+n+k) m n k} < \infty. \eeq
Please also note that for some trajectories Fourier coefficients may be zero for all possible virtual momenta. In this case $n_{pq}$ and $\kappa_{pq}$ do not receive growing with time contributions. An example of such a nonstationary motion can be found in Sec.~\ref{sec:synch}.

Let us emphasize that secular growth~\eqref{eq:n-two-fin} and~\eqref{eq:k-two-fin} has a clear physical origin. In a fully stationary situation energy conservation law forbids any kinetics (see App.~\ref{sec:artifacts}); however, the nonstationarity of the background violates this law\footnote{It is obvious that quantum scalar field and nonuniformly moving mirror form an open system. Therefore, external forces can pump in and out its energy. This is what we call a violation of the energy conservation law.} and allows usually forbidden processes to occur even at large evolution times. This violation is reflected in nonzero high order Fourier coefficients~\eqref{eq:Fourier}. Moreover, there are always contributions to~\eqref{eq:n-two} and~\eqref{eq:k-two} that do not depend on the integration times (or their transformed counterparts), because mode functions contain both incident and reflected waves. This results in the quadratic growth of quantum averages.

Thus, the problem of calculating loop corrections reduces to the problem of determining the Fourier coefficients of $G^{-1}(z)$. In general, these coefficients depend on the motion of the mirrors at intermediate times $0 < t < t_*$, so this problem is still challenging for arbitrary trajectories. However, for relatively simple functions $L(t)$ and $R(t)$ this approach is much more effective than the straightforward calculation of the integrals~\eqref{eq:n-two} and~\eqref{eq:k-two}. In the following subsections we illustrate this approach and estimate the leading contributions to $n_{pq}$ and $\kappa_{pq}$ for some particular functions $L(t)$ and $R(t)$.

\subsection{Simultaneous kicks}
\label{sec:sim}

In this subsection we consider the case of two simultaneous kicks ($t_L = t_R = 0$ in the notation~\eqref{eq:broken}):
\beq \label{eq:sim-tr}
L(t) = \beta t \theta(t), \quad R(t) = \Lambda + \beta t \theta(t) . \eeq
This type of the mirror motion is depicted on the Fig.~\ref{fig:two-mirrors}. For certainty we consider positive final velocities, $0 < \beta < 1$, although the discussion of this subsection is equally applicable to the case $-1 < \beta < 0$. Note that the distance between the mirrors in the observational reference frame is always fixed to be $R(t) - L(t) = \Lambda$. At the same time, the distance in the co-moving frame reduces from $\Lambda$ at the past infinity to $\Lambda \sqrt{1 - \beta^2}$ at the future infinity. 

Let us apply the geometrical method to find the modes for the trajectories~\eqref{eq:sim-tr}. First, the identity $F\left[ t - L(t) \right] = G\left[ t + L(t) \right]$ immediately implies $F(z) = G\left( \frac{1 + \beta}{1 - \beta} z \right)$ for $z > 0$, so it is sufficient to consider only the construction of $G(z)$. Second, moments of reflections are as follows:
\beq t_1 = \frac{z - \Lambda}{1 + \beta}, \quad \cdots, \quad t_{2k + 1} = t_1 - \frac{2 \Lambda k}{1 - \beta^2}, \quad t_{2k} = t_{2k - 1} - \frac{\Lambda}{1 - \beta}, \quad k = 1, 2, \cdots, n. \eeq
The total number of reflections is easily found using the periodical reflection pattern:
\beq n = \left\lceil \frac{1 - \beta}{2} \frac{z - \Lambda}{\Lambda} \right\rceil, \eeq
where $\lceil x \rceil$ denotes the least integer greater than or equal to $x$. Finally, $F$ and $G$ static regions in this picture correspond to the following values of $z > \Lambda$:
\beq \label{eq:sim-1} \begin{aligned}
\text{$F$ static region}: \quad &(n - 1) \frac{2 \Lambda}{1 - \beta} < z - \Lambda < n \frac{2 \Lambda}{1 - \beta} - \Lambda, \\
\text{$G$ static region}: \quad &n \frac{2 \Lambda}{1 - \beta} - \Lambda < z - \Lambda < n \frac{2 \Lambda}{1 - \beta}.
\end{aligned} \eeq
These identities imply that $G(z)$ is represented by the following piecewise linear function:
\beq \label{eq:sim-2}
G(z) = \begin{cases} \frac{1 - \beta}{1 + \beta} \frac{z}{\Lambda} + \frac{2 \beta n}{1 + \beta}, \quad & \text{for $F$ static region}, \\ \frac{z}{\Lambda} - \frac{2 \beta n}{1 - \beta}, \quad & \text{for $G$ static region}. \end{cases} \eeq
It is also convenient to introduce the new coordinate $\delta \equiv x - L(t)$, $0 < \delta < \Lambda$, which measures the distance to the left mirror. In these notations functions $G(t+x)$ and $F(t-x)$ are related by a simple shift:
\beq \label{eq:sim-3}
F(t, \delta) = F\left[ (1 - \beta) t - \delta \right] = G\left[ (1 + \beta) t - \frac{1 + \beta}{1 - \beta} \delta \right], \quad G(t, \delta) = G\left[ (1 + \beta) t + \delta \right]. \eeq
In this picture it is straightforward to see that functions $G$ and $F$ periodically grow with time, i.e. $G\left( t + \frac{2 \Lambda}{1 - \beta^2}, \delta \right) = G\left( t, \delta \right) + 2$ and $F\left( t + \frac{2 \Lambda}{1 - \beta^2}, \delta \right) = F\left( t, \delta \right) + 2$. When these functions are multiplied by $-i \pi n$ and substituted into the exponent of $e^x$, this shift yields a factor $e^{- 2 i \pi n} = 1$. Hence, the mode functions~\eqref{eq:mode-moore} are simply periodic with the period $\Delta t = \frac{2 \Lambda}{1 - \beta^2}$.

The explicit form of the mode functions straightforwardly follows from the identities~\eqref{eq:sim-1}, \eqref{eq:sim-2} and~\eqref{eq:sim-3}. The corresponding expressions can be found in the App.~\ref{sec:expl-modes}. These expressions are too bulky to integrate them explicitly for arbitrary evolution times. However, it is still possible to estimate the quantum averages~\eqref{eq:n-two} and~\eqref{eq:k-two} in the limits $T \ll \Lambda$ and $T \gg \Lambda$.

\subsubsection{Small evolution times and regularization}
\label{sec:sim-small}

Adapting the expressions from App.~\ref{sec:expl-modes} to the region $t < \frac{\Lambda}{1 + |\beta|}$ we obtain the following identities for the mode functions:
\beq \label{eq:sim-small}
g_n(t,x) = \begin{cases} \frac{i}{\sqrt{4 \pi n}} \left[ e^{-i \pi n \frac{t + x}{\Lambda}} - e^{-i \pi n \frac{1 + \beta}{1 - \beta} \frac{t - x}{\Lambda}} \right], \quad &\text{if} \quad \beta t < x < t, \\ \frac{i}{\sqrt{4 \pi n}} \left[ e^{-i \pi n \frac{t + x}{\Lambda}} - e^{-i \pi n \frac{t - x}{\Lambda}} \right], \quad &\text{if} \quad t < x < \Lambda - t, \\ \frac{i}{\sqrt{4 \pi n}} \left[ e^{-i \pi n \left( \frac{1 - \beta}{1 + \beta} \frac{t + x}{\Lambda} + \frac{2 \beta}{1 + \beta} \right)} - e^{-i \pi n \frac{t - x}{\Lambda}} \right], \quad &\text{if} \quad \Lambda - t < x < \Lambda + \beta t. \end{cases} \eeq
Note that in the $G$ static region these modes coincide with the modes for a single mirror (compare with~\cite{Alexeev}).

We would like to calculate integrals~\eqref{eq:n-two} and~\eqref{eq:k-two} in the limit $T \ll \Lambda$. Since the $\frac{T}{\Lambda}$ is a small parameter here, we can expand integrals into the Taylor series in $\frac{T}{\Lambda}$ and keep the leading order. Keeping in mind the stationary situation (App.~\ref{sec:artifacts}) we expect integrals $\int_0^T dt \int_{\beta t}^{\Lambda + \beta t} dx \, I_{p,m,n,k}(t, x)$ and $\int_0^T dt \int_{\beta t}^{\Lambda + \beta t} dx \, I^c_{p,m,n,k}(t,x)$ to grow linearly with $T$. At the same time, the areas of the leftmost ($\beta t < x < t$) and the rightmost ($\Lambda - t < x < \Lambda + \beta t$) regions in the definition~\eqref{eq:sim-small} are lesser than $(1 + |\beta|) T^2$. The contributions of these regions to the integrals are negligible in the limit in question. Hence, in the leading order the integrals of $I_{p,m,n,k}$ and $I^c_{p,m,n,k}$ coincide with the corresponding integrals in the stationary case:
\beq \begin{aligned}
I_{p,m,n,k}(T) &\approx \int_0^T dt \int_t^{\Lambda-t} dx \frac{e^{-\frac{i \pi t}{\Lambda}(p+m+n+k)}}{8 \pi^2 \sqrt{p m n k}} \sum_{\sigma_m, \sigma_n, \sigma_k = \pm 1} \sigma_m \sigma_n \sigma_k \cos \left[ \frac{\pi x}{\Lambda} (p + \sigma_m m + \sigma_n n + \sigma_k k) \right] \approx \\ &\approx \frac{T}{8 \pi^2 \sqrt{p m n k}} \sum_{\sigma_m, \sigma_n, \sigma_k = \pm 1} \sigma_m \sigma_n \sigma_k \Lambda \delta_{p + \sigma_m m + \sigma_n n + \sigma_k k, 0} + \mO(T^2).
\end{aligned} \eeq
The same conclusion applies to the integrals~\eqref{eq:n-two} and~\eqref{eq:k-two}. Hence:
\beq n_{pq} \approx n_{pq}^\text{stat} \sim (\lambda \Lambda T)^2, \quad \kappa_{pq} \approx \kappa_{pq}^\text{stat} \sim -(\lambda \Lambda T)^2. \eeq
In the limit of infinitely distant mirrors ($\Lambda \rightarrow \infty$ and $p = \frac{\pi p}{\Lambda} = \const$) this behavior can persist for a long time. However, this fake ``secular growth'' does not correspond to the change in the state of the system; in fact, it is just an artifact of the incorrect IR cut-off choice. This unphysical divergence can be cured by subtracting the corresponding quantities from the stationary theory:
\beq \label{eq:reg}
n_{pq}^\text{reg} \equiv n_{pq} - n_{pq}^\text{stat}, \quad \kappa_{pq}^\text{reg} \equiv \kappa_{pq} - \kappa_{pq}^\text{stat}. \eeq
Essentially, such a subtraction extends the integration intervals to large negative times, i.e. reproduces the calculations in the limit $t_0 \rightarrow -\infty$ instead of $t_0 = 0$ (compare with the calculations of two-loop corrections in~\cite{Alexeev}). Obviously, this way of regularization does not affect large times, $T \gg \Lambda$, because in this limit $n_{pq}^\text{stat} \rightarrow 0$ and $\kappa_{pq}^\text{stat} \rightarrow 0$ (see App.~\ref{sec:artifacts}). Hence, the regularization does not obscure the true secular growth which indicates the change in the state of the system and results in nonzero corrections to the stress-energy tensor.

Note that the analysis of this subsection does not depend on the form of trajectories $L(t)$ and $R(t)$ after $t = 0$ because for $T \ll \Lambda$ signals from the mirrors affect only regions with areas $\sim T^2$. Hence, the suggested regularization scheme can be applied to an arbitrary motion of the mirrors.

Finally, note that loop calculations of the Sec.~\ref{sec:non} and paper~\cite{Alexeev} implicitly assume exactly the same regularization for the quantum averages. Due to this reason we believe that these calculations correctly predict the behavior of $n_{pq}$ and $\kappa_{pq}$ in the single-mirror limit $\Lambda \rightarrow \infty$.

\subsubsection{Large evolution times}
\label{sec:sim-large}

Let us apply the machinery developed in the Sec.~\ref{sec:conformal} to estimate $n_{pq}$ and $\kappa_{pq}$ for large times, $T \gg \Lambda$. First, we rewrite the expression~\eqref{eq:sim-2} using Heaviside step functions:
\beq \begin{aligned}
G(z) = \theta(\Lambda - z) \frac{z}{\Lambda} + \sum_{n=1}^\infty \bigg\{ &\left[ \theta\left( \frac{z}{\Lambda} - 1 - \frac{2 (n - 1)}{1 - \beta} \right) - \theta\left( \frac{z}{\Lambda} - \frac{2 n}{1 - \beta} \right) \right] \left( \frac{1 - \beta}{1 + \beta} \frac{z}{\Lambda} + \frac{2 \beta n}{1 + \beta} \right) + \\ + &\left[ \theta\left(  \frac{z}{\Lambda} - \frac{2 n}{1 - \beta} \right) - \theta\left(  \frac{z}{\Lambda} - 1 - \frac{2 n}{1 - \beta} \right) \right] \left(  \frac{z}{\Lambda} - \frac{2 \beta n}{1 - \beta} \right) \bigg\}.
\end{aligned} \eeq
Having this representation it is straightforward to find the inverse function:
\beq \begin{aligned}
g(y) = \theta(1 - y) \Lambda y + \sum_{n=1}^\infty \bigg\{ &\bigg[ \theta\Big( y - (2 n - 1) \Big) - \theta\Big( y - 2 n \Big) \bigg] \left( \frac{1 + \beta}{1 - \beta} \Lambda y - \frac{2 \beta n \Lambda}{1 - \beta} \right) + \\ + &\bigg[ \theta\Big( y - 2 n \Big) - \theta\Big(  y - (2 n + 1) \Big) \bigg] \left( \Lambda y + \frac{2 \beta n \Lambda}{1 - \beta} \right) \bigg\},
\end{aligned} \eeq
and its derivative:
\beq g'(y) = \Lambda + \frac{2 \beta \Lambda}{1 - \beta} \sum_{n = 1}^\infty \Big[ \theta\Big( y - (2 n - 1) \Big) - \theta\Big( y - 2 n \Big) \Big]. \eeq
The corresponding Fourier coefficients are also easy to calculate:
\beq g_n = \frac{1}{2} \int_0^2 g'(y) e^{-i \pi n y} dy = -\frac{2 \beta \Lambda}{1 - \beta} \frac{1 - (-1)^n}{2 i \pi n}, \quad \text{for} \quad n \neq 0; \quad g_0 = \frac{\Lambda}{1 - \beta}. \eeq
Finally, using that in this case $t_* = 0$ we obtain the approximate expressions for the quantum averages:
\begin{align}
\label{eq:n-sim} n_{pq}(T) &\approx (\lambda \beta \Lambda T)^2 \times \frac{S_{p,q}}{32 \pi^6 \sqrt{p q}}, \\
\label{eq:k-sim} \kappa_{pq}(T) &\approx -(\lambda \beta \Lambda T)^2 \times \frac{S_{-p,q} + S_{p,-q}}{64 \pi^6 \sqrt{p q}},
\end{align}
where we introduced a short notation for the sum:
\beq S_{p,q} = \sum_{m,n,k=1}^\infty \frac{\left(1 - (-1)^{p + m + n + k} \right) \left(1 - (-1)^{q + m + n + k} \right)}{4 m n k (p + m + n + k) (q + m + n + k)}. \eeq
Note that $S_{p,q} = 0$ if $p$ and $q$ have different parity. However, $S_{pq}$ has nonzero nondiagonal terms that correspond to $p$ and $q$ of the same parity. This means that $n_{pq}$ and $\kappa_{pq}$ do not reduce to the diagonal form as it happens in the most cases (e.g. see~\cite{Akhmedov:dS, Akhmedov:Et, Akhmedov:Ex, Pavlenko}).

\begin{figure}[t]
\begin{minipage}[h]{0.49\linewidth}
\center{\includegraphics[width=1\linewidth]{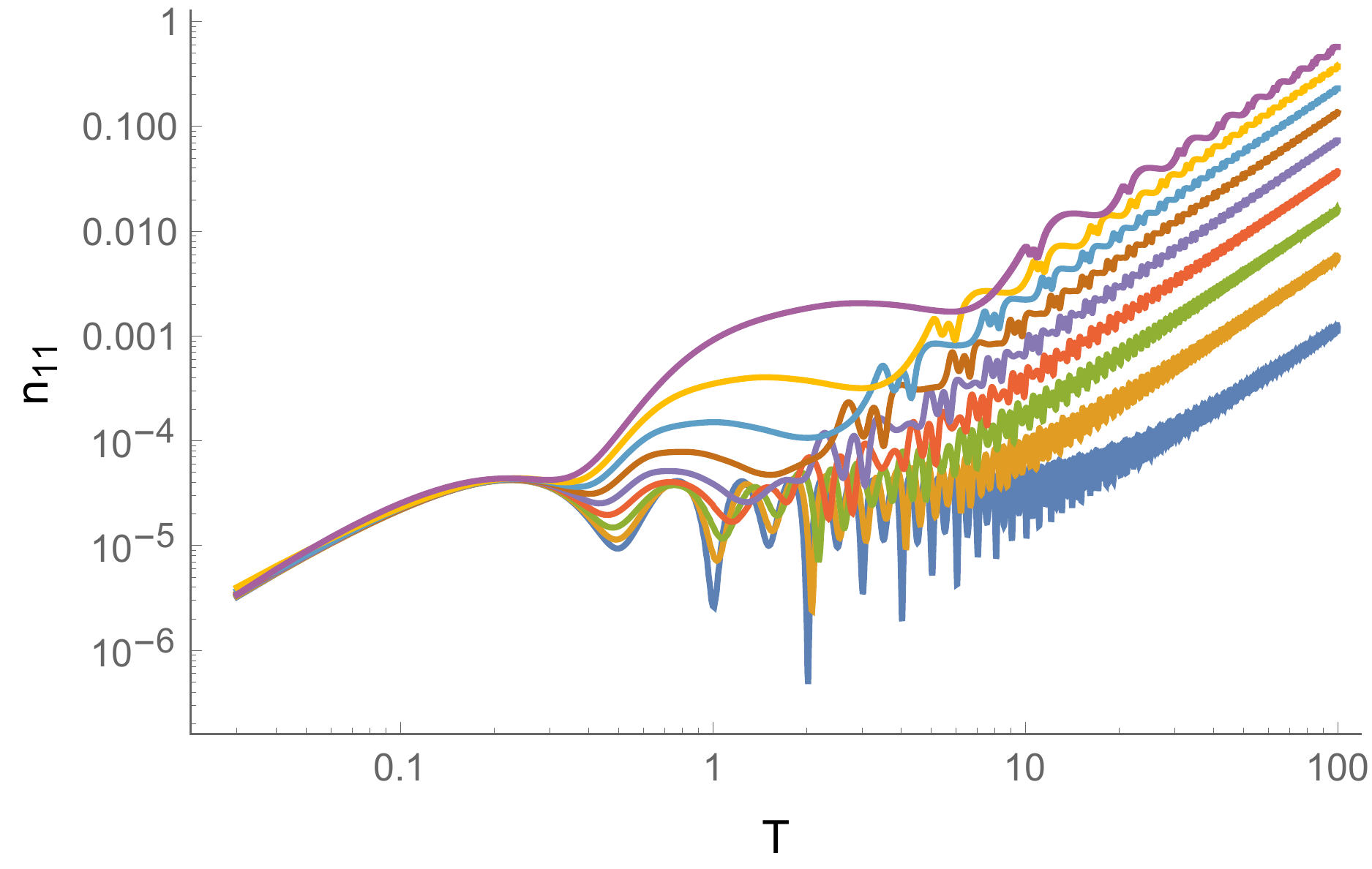} \\ a)}
\end{minipage}
\hfill
\begin{minipage}[h]{0.49\linewidth}
\center{\includegraphics[width=1\linewidth]{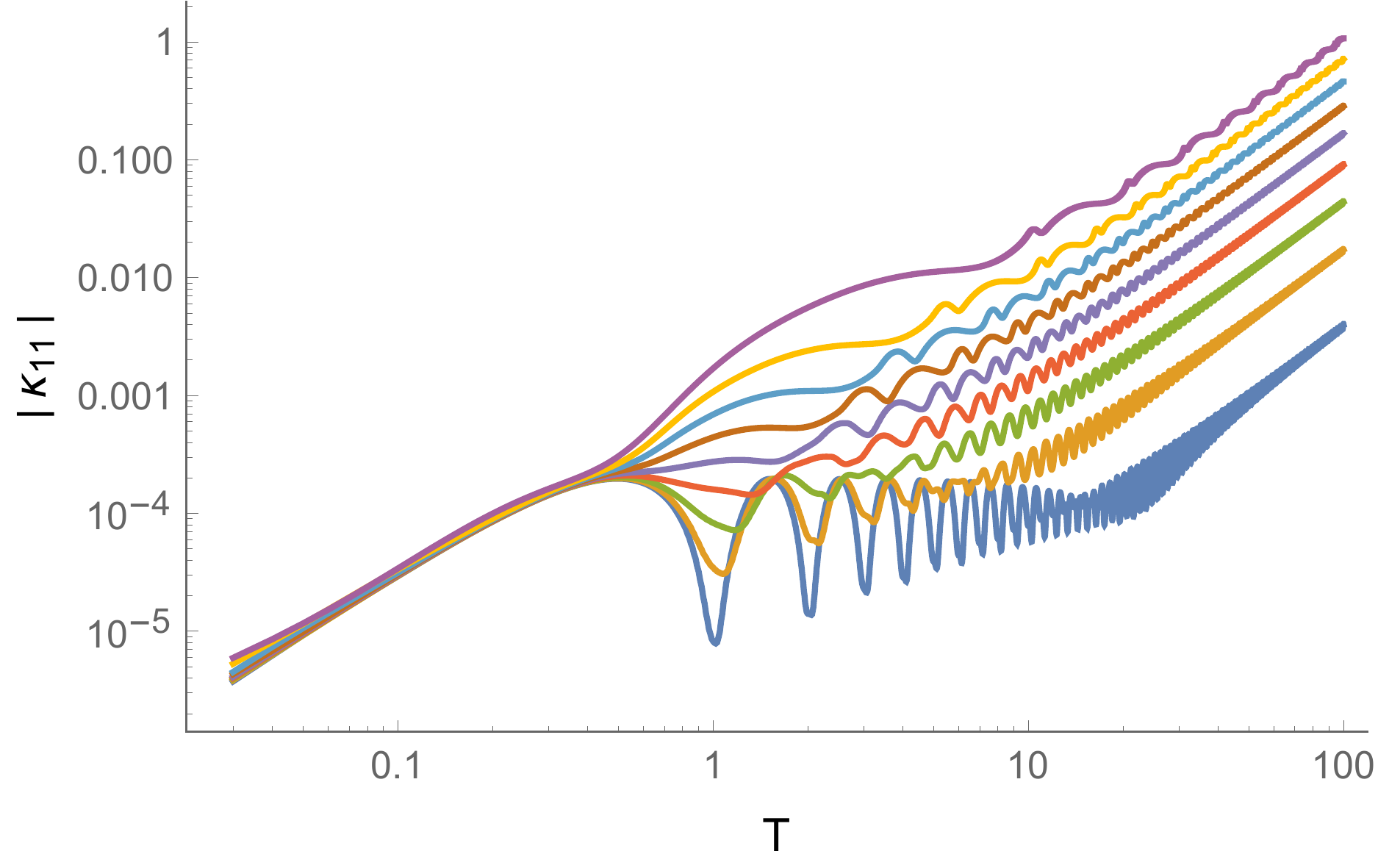} \\ b)}
\end{minipage}
\caption{Numerically estimated functions $n_{11}(T)$ (a) and $\left| \kappa_{11}(T) \right|$ (b). Different colors correspond to different mirror velocities $\beta$. The time $T$ is measured in units of $\Lambda$ and quantum averages are measured in units of $2 \lambda^2$. Note that the regularization~\eqref{eq:reg} was not applied.}
\label{fig:sim-graphs}
\end{figure}

Thus, in the case of simultaneous kicks~\eqref{eq:sim-tr} quantum averages quadratically grow both with small ($T \ll \Lambda$) and large ($T \gg \Lambda$) evolution times. The regularization~\eqref{eq:reg} eliminates the meaningless growth at small times and does not affect the meaningful growth at large times. In the intermediate region these asymptotics are connected by a smooth function which can be calculated numerically (Fig.~\ref{fig:sim-graphs}).

Finally, note that in the limit $\Lambda \rightarrow \infty$ and $p = \frac{\pi p}{\Lambda} = \const$ two-mirror problem qualitatively reproduces the single-mirror problem considered in~\cite{Alexeev}. In both these cases quantum averages quadratically grow with evolution time, although the velocity-dependent prefactors of these growths are slightly different: $n_{pq} \sim \frac{\beta^2}{(1 + \beta)^2} \frac{S_{pq}}{\sqrt{pq}} (\lambda T)^2$ in the one-mirror case and $n_{pq} \sim \beta^2 \frac{S_{pq}}{\sqrt{pq}} (\lambda T)^2$ in the two-mirror case. For small final velocities, $|\beta| \ll 1$, this difference is negligible, which extends the correspondence to a quantitative level.

\subsection{Synchronized kicks}
\label{sec:synch}

The other notable option to adjust mirror trajectories is to consider synchronized kicks\footnote{One can also synchronize kicks along the $u = \const$ ray, i.e. set $t_L + \Lambda = t_R$. This motion has the same properties as the motion~\eqref{eq:synch-tr}.} connected by a null line ($t_L = t_R + \Lambda$ in the notation~\eqref{eq:broken}):
\beq \label{eq:synch-tr}
L(t) = \beta (t - \Lambda) \theta(t - \Lambda), \quad R(t) = \Lambda + \beta t \theta(t). \eeq
In this case $F$ and $G$ static regions coincide, which significantly simplifies the derivation of $G(z)$:
\beq G(z) = \frac{1 - \beta}{1 + \beta} \frac{z}{\Lambda} + \frac{2 \beta}{1 + \beta}, \quad \text{for} \quad z > \Lambda. \eeq
Furthermore, it is straightforward to show that $F(z) = \frac{z}{\Lambda}$ for all $z$.

We would like to estimate the large-time asymptotics of functions $n_{pq}(T)$ and $\kappa_{pq}(T)$ which do not depend on the low-time behavior of the mode functions. Hence, for our purposes it is sufficient to consider only the region $t > \frac{2 \Lambda}{1 - \beta}$ where the mode functions~\eqref{eq:mode-moore} acquire the following form:
\beq g_n(t,x) = \frac{i}{\sqrt{4 \pi n}} e^{- i \pi n (1 - \beta) \frac{t}{\Lambda} - i \pi n \beta} \left[ e^{- i \pi n \frac{1 - \beta}{1 + \beta} \frac{\delta}{\Lambda}} - e^{i \pi n \frac{\delta}{\Lambda}} \right], \eeq
where $\delta \equiv x - L(t)$ and $0 < \delta < (1 + \beta) \Lambda$ for all $t > \frac{2 \Lambda}{1 - \beta}$. Substituting these functions into integrals~\eqref{eq:n-two} and~\eqref{eq:k-two} we obtain that neither $n_{pq}$ nor $\kappa_{pq}$ grows in the limit $T \rightarrow \infty$:
\begin{align}
n_{pq}(T) &\approx 2 \lambda^2 \int_{\frac{2\Lambda}{1 - \beta}}^T dt_1 \int_{\frac{2\Lambda}{1 - \beta}}^T dt_2 \sum_{m,n,k = 1}^\infty e^{- i \pi (1 - \beta) (p + m + n + k) \frac{t_1}{\Lambda}} e^{i \pi (1 - \beta) (q + m + n + k) \frac{t_2}{\Lambda}} C(p,q,m,n,k) \sim \nonumber \\ &\sim \lambda^2 \Lambda^4, \label{eq:n-synch} \\
\kappa_{pq}(T) &\approx 2 \lambda^2 \int_{\frac{2\Lambda}{1 - \beta}}^T dt_1 \int_{\frac{2\Lambda}{1 - \beta}}^{t_1} dt_2 \sum_{m,n,k = 1}^\infty e^{i \pi (1 - \beta) (p + m + n + k) \frac{t_1}{\Lambda}} e^{i \pi (1 - \beta) (q + m + n + k) \frac{t_2}{\Lambda}} D(p,q,m,n,k) \sim \nonumber \\ &\sim \lambda^2 \Lambda^4. \label{eq:k-synch}
\end{align}
Here we have introduced functions $C$ and $D$ that depend on momenta but do not depend on times $t_1$ and $t_2$:
\begin{align}
C(p,q,m,n,k) &= \int_0^\Lambda d\delta_1 \int_0^\Lambda d\delta_2 \, \tilde{g}_{p,1} \tilde{g}_{q,2}^* \tilde{g}_{m,1} \tilde{g}_{n,1} \tilde{g}_{k,1} \tilde{g}_{m,2}^* \tilde{g}_{n,2}^* \tilde{g}_{k,2}^*, \\
D(p,q,m,n,k) &= \int_0^\Lambda d\delta_1 \int_0^\Lambda d\delta_2 \left( \tilde{g}_{p,1}^* \tilde{g}_{q,2}^* + \tilde{g}_{q,1}^* \tilde{g}_{p,2}^* \right) \tilde{g}_{m,1} \tilde{g}_{n,1} \tilde{g}_{k,1} \tilde{g}_{m,2}^* \tilde{g}_{n,2}^* \tilde{g}_{k,2}^*,
\end{align}
where $\tilde{g}_{p,n} = \frac{i}{\sqrt{4 \pi n}} e^{- i \pi n \beta} \left[ e^{- i \pi n \frac{1 - \beta}{1 + \beta} \frac{\delta}{\Lambda}} - e^{i \pi n \frac{\delta}{\Lambda}} \right]$. For the latter identities in~\eqref{eq:n-synch} and~\eqref{eq:k-synch} we used that time integrals are bounded (there are no singular contributions because the arguments of the exponential functions are never zero) and the sums over the virtual momenta are convergent:
\begin{gather*}
\left| C(p,q,m,n,k) \right| \sim \left| D(p,q,m,n,k) \right| \lesssim \frac{1}{\pi^4 \sqrt{p q}} \frac{\Lambda^2}{m n k}, \quad \text{hence}, \\
\sum_{m,n,k=1}^\infty \frac{\Lambda^2 \times \left[ C \; \text{or} \; D \right]}{(p+m+n+k) (q+m+n+k)} \sim \sum_{m,n,k=1}^\infty \frac{\Lambda^4}{(p+m+n+k) (q+m+n+k) m n k} < \infty.
\end{gather*}
Note that the dimensionless coefficients of proportionality in identities~\eqref{eq:n-synch} and~\eqref{eq:k-synch} include the velocity $\beta$.

One can also come to the same conclusion using the approach of the Sec.~\ref{sec:conformal}. Indeed, in this case we have $g(y) = \frac{1 + \beta}{1 - \beta} \Lambda y - \frac{2 \beta \Lambda}{1 - \beta}$ and $g'(y) = \frac{1 + \beta}{1 - \beta} \Lambda$ for $y > y_* = 2$. Therefore, the only nonzero coefficient of the Fourier expansion is $g_0 = \frac{1 + \beta}{1 - \beta} \Lambda$. At the same time, in our case all frequencies are positive, i.e. $p + m + n + k > 0$ for all $p,m,n,k$. Hence, it is impossible to get nonzero contributions to the sum over virtual momenta in~\eqref{eq:n-two-fin} and~\eqref{eq:k-two-fin}. This implies that $n_{pq}$ and $\kappa_{pq}$ do not receive growing with time loop corrections.

Note that the difference between synchronized ($t_L = t_R \pm \Lambda$) and unsynchronized ($t_L \neq t_R \pm \Lambda$, e.g. $t_L = t_R$) kicks appears already at the tree level~\cite{Davies:1976}. Namely, it can be shown that in the case of synchronized kicks the tree-level stress-energy tensor receives nonzero contributions only at the intermediate times, $0 < t < \Lambda$ (the setup discussed in the paper~\cite{Davies:1976} is more complex than~\eqref{eq:synch-tr} but the calculations in these two cases are essentially the same). At the same time, in the unsynchronized case the signal and the radiation pulse from the first kick ``misses'' the second kick and bounces back and forth between the mirrors indefinitely. In the paper~\cite{Davies:1976} this tree-level effect was interpreted as evidence of particle creation. In the former case this process is temporary (the radiation emitted during the first kick is completely absorbed during the second kick), and in the latter case it is permanent.

It is noteworthy that this difference persists at the loop level: if the kicks are synchronized, then both tree-level and IR two-loop-level contributions to the stress-energy tensor are negligible, whereas in the opposite case both contributions are significant. Recall that loop corrections to the stress-energy tensor are derived from the quantum averages using~\eqref{eq:T}. This difference also implies that ground state of the system behaves differently in the cases of synchronized and unsynchronized kicks. 

From the point of view of Secs.~\ref{sec:geometry} and~\ref{sec:conformal} the difference between the cases of synchronized and unsynchronized kicks can be easily explained as follows. First, in the case of synchronized kicks the number of reflections from the left and the right mirror always differ by one. Second, the synchronization requirement implies that $R(t_{2i - 1}) - L(t_{2i}) = L$ for $i = 1, \cdots, n - 1$. Hence, only the first reflection point makes a nontrivial contribution to the identity~\eqref{eq:G}. Finally, recall that we consider asymptotically stationary motions, i.e. we assume that $\dot{L}(t) = \dot{R}(t) = \beta$ for $t > t_*$. Together these observations imply that function $G(z)$ is purely linear for $z > z_*$ and $g'(y)$ is constant for $y > y_*$. Therefore, the high order Fourier coefficients~\eqref{eq:Fourier} are zero and integrals~\eqref{eq:n-two-fin} and~\eqref{eq:k-two-fin} cannot receive secularly growing loop corrections. The same reasoning also works for complex synchronized motions, e.g. the case where the second kick is applied after several reflections of the first kick.

In other words, the fine-tuning of kicks ensures the energy conservation law in the infinite future which, in turn, leads to the zero collision integral (compare with App.~\ref{sec:artifacts}). Note that for a finite-duration nonstationary motion synchronization must be performed during the entire motion period. At the same time, without any synchronization this argumentation does not work, energy conservation is violated and secular growth is possible.

\subsection{Resonant cavity}
\label{sec:cavity}

Finally, let us apply the approach of Sec.~\ref{sec:conformal} to a one-dimensional resonantly oscillating cavity:
\beq \begin{aligned}
L(t > 0) &= \epsilon \Lambda \sin \left(\frac{s \pi t}{\Lambda} \right), \\
R(t > 0) &= \Lambda + \epsilon \Lambda \sin\left( \frac{s \pi t}{\Lambda} + \varphi \right) - \epsilon \Lambda \sin \varphi,
\end{aligned} \eeq
where $\epsilon \ll 1$ is the small parameter, $s \in \mathbb{N}$ defines the frequency of oscillations and $\varphi$ is the dephasing angle. For illustrative purposes we set $\varphi = 0$ (this describes the cavity oscillating as a whole) and $s = 2$. In this case functions $G(z)$ and $F(z)$ have the following form~\cite{Dalvit:1997, Dalvit:1998}:
\beq \label{eq:GF-cavity} \begin{aligned}
G(z) &= \frac{z}{\Lambda} - 2 \epsilon \sin \frac{2 \pi z}{\Lambda} \sum_{n = 1}^\infty \left[ \theta\left( \frac{z}{\Lambda} - (2 n - 1) \right) - \theta\left( \frac{z}{\Lambda} - 2 n \right) \right] + \mO(\epsilon^2), \\
F(z) &= \frac{z}{\Lambda} + 2 \epsilon \sin \frac{2 \pi z}{\Lambda} \sum_{n = 0}^\infty \left[ \theta\left( \frac{z}{\Lambda} - 2 n \right) - \theta\left( \frac{z}{\Lambda} - (2 n + 1) \right) \right] + \mO(\epsilon^2).
\end{aligned} \eeq
This solution is valid even for relatively large arguments\footnote{Note that the naive approach based on a perturbative expansion in $\epsilon$ is applicable only for $z \ll \Lambda / \epsilon$, because at larger arguments identities~\eqref{eq:GF-cavity} receive secularly growing corrections of the form $\epsilon^n t^m$. The leading corrections of the form $\epsilon^n t^n$ can be resummed using a renormalization group technique discussed in~\cite{Dalvit:1997, Dalvit:1998}. However, for the case $s=2$ this resummation does not result in new corrections to the naive formula, so it can be simply extended to $\Lambda / \epsilon \ll z \ll \Lambda/\epsilon^2$.}, $\Lambda / \epsilon \ll z \ll \Lambda/\epsilon^2$. Now one can see that functions $G(z)$ and $F(z)$ are approximately periodic. Hence, the approach of Sec.~\ref{sec:conformal} is applicable even though the mirrors motion is not uniform at large times. The inverse functions in this case are straighforwardly determined up to the same order in $\epsilon$:
\beq \begin{aligned}
\frac{g(y)}{\Lambda} &= y + 2 \epsilon \sin\left( 2 \pi y \right) \sum_{n = 1}^\infty \left[ \theta\left( y - (2 n - 1) \right) - \theta\left( y - 2 n \right) \right] + \mO(\epsilon^2), \\
\frac{f(y)}{\Lambda} &= y - 2 \epsilon \sin\left( 2 \pi y \right) \sum_{n = 0}^\infty \left[ \theta\left( y - 2 n \right) - \theta\left( y - (2 n + 1) \right) \right] + \mO(\epsilon^2).
\end{aligned} \eeq
This approximation is valid for $y \ll 1/\epsilon^2$. Now it is easy to see that high Fourier coefficients are nonzero:
\beq g_n = \epsilon \Lambda \frac{- 4 i n}{n^2 - 4} \frac{1 - (-1)^n}{2}, \quad \text{for} \quad n \neq -2, 0, 2; \quad g_0 = \frac{1}{2} \Lambda, \quad \text{and} \quad g_{\pm 2} = \pi \epsilon \Lambda. \eeq
Therefore, quantum averages receive secularly growing loop corrections which are significant at relatively large times, $\Lambda / \epsilon \ll t \ll \Lambda / \epsilon^2$:
\begin{align}
n_{pq}(T) &\approx (\lambda \epsilon \Lambda T)^2 \frac{S_{p,q}}{8 \pi^4 \sqrt{pq}}, \\
\kappa_{pq}(T) &\approx -(\lambda \epsilon \Lambda T)^2 \frac{S_{-p,q} + S_{p,-q}}{16 \pi^4 \sqrt{pq}}.
\end{align}
Here we neglected the subleading $\mO(\epsilon^2)$ and oscillating contributions. Also we introduced a short notation for the sum over virtual momenta:
\beq S_{p,q} = \sum_{m,n,k=1}^\infty \frac{\left( 1 - (-1)^{p + m + n + k} \right) \left( 1 - (-1)^{q + m + n + k} \right)}{4 m n k} \frac{p + m + n + k}{(p + m + n + k)^2 - 4} \frac{q + m + n + k}{(q + m + n + k)^2 - 4}. \eeq
Thus, we have shown that a self-interacting massless scalar field on the background of resonantly oscillating mirrors receives secularly growing loop corrections to the quantum averages. This indicates the change in the ground state of the system. Also this means that loop corrections may affect particle production in a resonant cavity. However, we emphasize that the final conclusion about the destiny of the ground state and stress-energy flux can be made only after the resummation of the leading corrections from all loops.

\section{Single nonideal mirror}
\label{sec:non}

In this section we consider loop corrections to the DCE on a single nonideal mirror background. First of all, we discuss the quantization of a free two-dimensional massless scalar field on such a background. We model the mirror with the delta-functional potential. Using the established mode decomposition we calculate loop corrections to the energy level density and anomalous quantum average. For simplicity we assume that mirror moves along a ``broken'' trajectory~\eqref{eq:broken}.

\subsection{Free field quantization}
\label{sec:non-tree}

Consider a free two-dimensional massless scalar field with the delta-potential background:
\beq \label{eq:non-free}
S = \int d^2 x \left[ \frac{1}{2}( \pd_\mu \phi)^2 - \frac{\alpha}{2} \delta\left( \frac{x - x(t)}{\sqrt{1 - \beta^2(t)}} \right) \phi^2 \right], \eeq
where function $x(t)$ determines the position of the mirror at the moment $t$, $\beta(t) \equiv \frac{d x(t)}{dt}$ is the velocity of the mirror ($|\beta(t)| < 1$ for all $t$) and $\alpha$ is the coefficient that controls the ``ideality'' of the mirror (the mirror is perfectly reflective when $\alpha \rightarrow \infty$ and perfectly transparent when $\alpha = 0$). The quantized field can be represented by the mode decomposition:
\beq \label{eq:decomposition}
\phi(t,x) = \int_{-\infty}^\infty \frac{dp}{2 \pi} \left[ a_p g_p(t,x) + a_p^\dagger g_p^*(t,x) \right]. \eeq
Here $a_p^\dagger$ and $a_p$ are creation and annihilation operators that obey the standard commutation relations, $\left[ a_p, a_q^\dagger \right] = 2 \pi \delta(p - q)$; mode functions $g_p$ solve the corresponding equation of motion:
\beq \label{eq:free-non}
\left[ \pd_\mu \pd^\mu + \alpha \delta\left( \frac{x - x(t)}{\sqrt{1 - \beta^2(t)}} \right) \right] g_p(t,x) = 0, \eeq
and satisfy orthonormality conditions:
\beq \label{eq:ortho}
(g_p, g_q) = \delta(p - q), \quad (g_p, g_q^*) = 0, \eeq
w.r.t. the Klein-Gordon inner product~\cite{Birrell, DeWitt}:
\beq (f, h) \equiv -i \int_{-\infty}^\infty dx \Big[ f(t,x) \pd_t h^*(t,x) - h^*(t,x) \pd_t f(t,x) \Big]. \eeq
Note that relations~\eqref{eq:decomposition} and~\eqref{eq:ortho} automatically imply the canonical equal-time commutation relations $\left[ \phi(t,x), \pd_t \phi(t,y) \right] = i \delta(x - y)$. Also note that in the ideal-mirror case ($\alpha \rightarrow \infty$) the r.h.s of this identity contains additional boundary terms because in this case modes have improper UV behavior (e.g. see~\cite{Alexeev, Astrakhantsev}). As we will see below, on the nonideal mirror background reflected waves are negligible in the UV limit, i.e. high-energy modes behave as simple plane waves. Hence, this theory does not suffer from the problems of the ideal mirror theory.  

In the paper~\cite{Nicolaevici} it was shown that mode functions satisfying these conditions can be represented as the sum of reflected and transmitted waves:
\beq \label{eq:mode} \begin{aligned}
g_p(t,x) &= \theta(p) \left[ \theta \left( x(t) - x \right) \left( \frac{e^{- i \omega u}}{\sqrt{2 \omega}} - R_{\omega}^L(v) \frac{e^{- i \omega f(v)}}{\sqrt{2 \omega}} \right) + \theta\left( x - x(t) \right) T_{\omega}^R(u) \frac{e^{- i \omega u}}{\sqrt{2 \omega}} \right] + \\ &+  \theta(-p) \left[ \theta \left( x - x(t) \right) \left( \frac{e^{- i \omega v}}{\sqrt{2 \omega}} - R_{\omega}^R(u) \frac{e^{- i \omega g(u)}}{\sqrt{2 \omega}} \right) + \theta\left( x(t) - x \right) T_{\omega}^L(v) \frac{e^{- i \omega v}}{\sqrt{2 \omega}} \right], 
\end{aligned} \eeq
where $\omega \equiv |p|$, $u \equiv t - x$ and $v = t + x$ are light-cone coordinates, and functions $f$, $g$ are chosen such that identities $f(v) = u$, $g(u) = v$ are satisfied when the point $(u, v)$ moves along the trajectory of the mirror (i.e. when $u = t - x(t)$, $v = t + x(t)$). In these notations positive(negative)-momentum modes correspond to the right(left)-moving waves. The reflection and transmission coefficients on the mirror are fixed by the stitching conditions which imply the following expressions (here we additionally assume that the velocity of the mirror is uniform in the infinite past):
\beq \label{eq:RT} \begin{aligned}
R_\omega^R(\tau) &= \frac{\alpha}{2} \int_{-\infty}^\tau d\tau' \exp\left[ i \omega \left( v(\tau) - v(\tau') \right) - \frac{\alpha}{2} (\tau - \tau') \right], & \quad T_\omega^L(\tau) &= 1 - R_\omega^R(\tau), \\
R_\omega^L(\tau) &= \frac{\alpha}{2} \int_{-\infty}^\tau d\tau' \exp\left[ i \omega \left( u(\tau) - u(\tau') \right) - \frac{\alpha}{2} (\tau - \tau') \right], & \quad T_\omega^R(\tau) &= 1 - R_\omega^L(\tau),
\end{aligned} \eeq
where $\tau(t) = \tau_0 + \int_{t_0}^t dt \sqrt{1 - \beta^2(t)}$ is the proper time of the mirror and functions $u(\tau)$, $v(\tau)$ denote the corresponding coordinates on the mirror. Also we treated coefficients $R$ and $T$ as functions of $\tau$ via $u = u(\tau)$, $v = v(\tau)$. Note that $u(\tau)$ and $v(\tau)$ are invertible functions if the trajectory of the mirror is time-like. Hence, we can define proper times $\tau_u$ and $\tau_v$ such that $u(\tau_u) = u$ and $v(\tau_v) = v$. These are the proper times of projections of the point $(u,v)$ onto the mirror along the lines $u = \const$ and $v = \const$, respectively. Using this notation we can restore coefficients $R$ and $T$ for an arbitrary space-time point: $R_\omega^L(v) = R_\omega^L(\tau_v)$, $R_\omega^R(u) = R_\omega^R(\tau_u)$, $T_\omega^L(v) = T_\omega^L(\tau_v)$, $T_\omega^R(u) = T_\omega^R(\tau_u)$.

Note that in the limit $\omega \gg \alpha$ reflection coefficients tend to zero due to fast oscillations of integrands in~\eqref{eq:RT}. Hence, in the UV region the reflected waves are negligible, i.e. modes~\eqref{eq:mode} reduce to simple plane waves. This means that in the UV region theory with a nonideal mirror reduces to a standard 2D theory in empty space, so the UV renormalizations can be performed in a standard way.

For an arbitrary trajectory coefficients~\eqref{eq:RT} are very hard to find explicitly, so we need to make an approximation to keep the calculations feasible. First, note that for a uniform trajectory $x(t) = \beta t$ integrals in~\eqref{eq:RT} can be explicitly taken:
\beq \label{eq:RT-uniform}
R_\omega^{R,L}(\tau) = \frac{\alpha/2}{\alpha/2 - i \omega D_\beta^\pm}, \quad \text{where} \quad D_\beta^\pm = \sqrt{\frac{1 \pm \beta}{1 \mp \beta}}. \eeq
Here $D_\beta^\pm$ are the Doppler factors for the right and left incident waves, so this formula has a transparent interpretation in terms of Doppler shifts. Second, integrals~\eqref{eq:RT} are predominantly gained on the interval $0 < \tau - \tau' < \frac{1}{\alpha}$ due to the exponential decay of the integrand. Now note that trajectory $x(t)$ in the integrals~\eqref{eq:RT} can be approximated by a line at times $0 < \tau - \tau' \ll \frac{\alpha v'(\tau)}{v''(\tau)} \approx \frac{\alpha}{\gamma^3(t) |\ddot{x}(t)|}$. Hence, for a trajectory with relatively small proper acceleration, $\left| w(t) \right| = \gamma^3(t) \left| \ddot{x}(t) \right| \ll \alpha$, reflection coefficients can be approximated by~\eqref{eq:RT-uniform}:
\beq \label{eq:RT-app}
R_\omega^{R,L}(\tau) \approx \frac{\alpha/2}{\alpha/2 - i \omega D_\beta^\pm(\tau)} + \mO\left( \frac{w}{\alpha} \right), \quad \text{where} \quad D_\beta^\pm(\tau) = \sqrt{\frac{1 \pm \beta(\tau)}{1 \mp \beta(\tau)}}. \eeq
As was discussed in the Sec.~\ref{sec:picture}, a realistic coefficient is $\alpha \sim 10^{1 \div 5}$ cm$^{-1}$. This gives an estimated threshold acceleration $w \sim c^2 \alpha \sim 10^{20 \div 24}$ m/s$^2$. Therefore, for the most practical situations corrections to~\eqref{eq:RT-uniform} can be neglected (although some experiments with plasma acceleration can achieve almost as large $w$~\cite{Leemans}).

\subsection{Loop corrections}
\label{sec:non-loop}

In this subsection we use mode functions~\eqref{eq:mode} with the approximate reflection and transmission coefficients~\eqref{eq:RT-app} to calculate loop corrections to the energy level density and anomalous quantum average. We consider corrections generated by the $\lambda \phi^4$ nonlinearity:
\beq S = \int d^2 x \left[ \frac{1}{2}( \pd_\mu \phi)^2 - \frac{\alpha}{2} \delta\left( \frac{x - x(t)}{\sqrt{1 - \beta^2(t)}} \right) \phi^2 - \frac{\lambda}{4} \phi^4 \right]. \eeq
We would like to single out only the leading quantum corrections, so we work in the limit of small coupling constants and large evolution times, $\lambda \rightarrow 0$, $T \rightarrow \infty$, $\lambda T = \const$. We remind that in this limit the leading corrections to $n_{pq}$ and $\kappa_{pq}$ are given by eqs.~\eqref{eq:n} and~\eqref{eq:k}, respectively.

Similarly to the case of two ideal mirrors (Sec.~\ref{sec:conformal}), energy level density~\eqref{eq:n} can be represented as the product of two integrals:
\beq \label{npq}
n_{pq} \approx 2 \lambda^2 \int \frac{dk_1 dk_2 dk_3}{(2 \pi)^3} I_p(T) I_q^*(T),  \eeq
where introduced a short notation for the integral:
\beq I_p(T) = \int_{t_0}^T dt_1 \int_{-\infty}^\infty dx_1 \, g_{p,1} g_{k_1,1} g_{k_2,1} g_{k_3,1}, \quad \text{where} \quad g_{p,n} \equiv g_p(t_n, x_n). \eeq
We would like to single out the secularly growing parts of the integral $I_p(T)$. Such secular growth appears only when the integrand reduces to a product of the functions which depend on the same argument and has the same support. In the opposite case, i.e. when the integrand depends on both $u$ and $v$, the resulting integral oscillates or decays when $T \rightarrow \infty$. Therefore, such terms do not contribute to the secular growth and can be neglected. Also recall that $\mO\left( w/\alpha \right)$ terms are practically negligible.

In the mentioned approximation we straightforwardly obtain that $I_p(T)$ linearly grows with time:
\beq I_p(T) = \frac{i T}{4 \sqrt{|p k_1 k_2 k_3|} \left( |p| + |k_1| + |k_2| + |k_3| \right)} g(p, k_1, k_2, k_3). \eeq
Here we have separated the universal prefactor and the variable term which consists of 8 different products of the mode parts:
\beq \label{I_1} \begin{aligned}
g(p, k_1, k_2, k_3) &= \Big[ \theta(p) \theta(k_1) \theta(k_2) \theta(k_3) + \theta(-p) \theta(-k_1) \theta(-k_2) \theta(-k_3) \Big] \times \\
&\times \bigg[  T_{|p|}T_{|k_3|}T_{|k_2|}T_{|k_1|}+R_{|p|}R_{|k_3|}R_{|k_2|}R_{|k_1|} -\\
&-\left (T^{-c}_{|p|}T^{-c}_{|k_3|}T^{-c}_{|k_2|}T^{-c}_{|k_1|}+R^{-c}_{|p|}R^{-c}_{|k_3|}R^{-c}_{|k_2|}R^{-c}_{|k_1|}[D_\beta^c]^2\right )e^{-is(T-cx(T))} + \\
&+\int_0^{T-cx(T)} d[-isu] T^{-c}_{|p|}(u)T^{-c}_{|k_3|}(u)T^{-c}_{|k_2|}(u)T^{-c}_{|k_1|}(u)e^{-isu}+ \\
&+\int_0^{T+cx(T)} d[-isv]R^{-c}_{|p|}(v)R^{-c}_{|k_3|}(v)R^{-c}_{|k_2|}(v)R^{-c}_{|k_1|}(v) e^{-is(2t_v-v)}\bigg ] + \sum_{\{p\}} J^{\{p\}}_{u,v},
\end{aligned} \eeq
where $c=\sgn(p)$, $s=|p|+|k_1|+|k_2|+|k_3|$, times $t_u$ and $t_v$ solve the equations $u=t_u-x(t_u)$ and $v=t_v+x(t_v)$. Also we have introduced a short notation for the transmission and reflection coefficients of the stationary mirror: $T_\omega = \frac{2 i \omega}{2 i \omega - \alpha}$, $R_\omega = \frac{\alpha}{2 i \omega - \alpha}$, and moving mirror: $T^c_\omega(u) = \frac{2 i \omega D_\beta^c(u)}{2 i \omega D_\beta^c(u) - \alpha}$, $R^c_\omega(u) = \frac{\alpha}{2 i \omega D_\beta^c(u) - \alpha}$. These coefficients come from the $t < 0$ and $t > 0$ parts of the mirror trajectory, respectively. For brevity we have presented only four secularly growing terms which correspond to a single combination of the mode parts; the others are denoted as $\sum_{\{p\}} J^{\{p\}}_{u,v}$ and have the same structure. The explicit form of these terms can be found in the App.~\ref{sec:loop_cor}.

Analytic calculation of the integrals in~\eqref{I_1} cannot be performed for an arbitrary mirror trajectory. However, in the limit of a ``broken'' trajectory~\eqref{eq:broken} oscillating expressions reduce each other. Roughly speaking, these oscillating parts are non-zero only at the space-time regions that are causally connected with the segment of the accelerated motion of the mirror. For a ``broken'' trajectory this segment degenerates into a single dot, so the leading order term of $g(p, k_1, k_2, k_3)$ is simplified:
\beq \begin{aligned}
g(p, k_1, k_2, k_3) &= \left[ \theta(p) \theta(k_1) \theta(k_2) \theta(k_3) + \theta(-p) \theta(-k_1) \theta(-k_2) \theta(-k_3) \right] \times \\
&\times \Big[  T_{|p|} T_{|k_3|} T_{|k_2|} T_{|k_1|} + R_{|p|} R_{|k_3|} R_{|k_2|} R_{|k_1|} - \\ &- T^{-c}_{|p|} T^{-c}_{|k_3|} T^{-c}_{|k_2|} T^{-c}_{|k_1|} - R^{-c}_{|p|} R^{-c}_{|k_3|} R^{-c}_{|k_2|} R^{-c}_{|k_1|} ( D_\beta^c )^2 \Big] + \sum_{\{p\}} J^{\{p\}}_{u,v}. \end{aligned} \eeq
Here once again $J^{\{p\}}_{u,v}$ is the sum of all terms proportional to the mixed products of transmission and reflection coefficients. Substituting the calculated expression for the integral $I_p(T)$ into the energy level density~\eqref{npq}, we obtain:
\beq \label{eq:npq}
n_{pq} = \frac{(\lambda T)^2}{8 \sqrt{|p q|}} \int \frac{dk_1 dk_2 dk_3}{(2\pi)^3 |k_1k_2k_3|} \frac{g(p,k_1,k_2,k_3) g^*(q,k_1,k_2,k_3)}{\left( |p| + |k_1| + |k_2| + |k_3| \right) \left( |q| + |k_1| + |k_2| + |k_3| \right)} + \mO(\lambda^2 T). \eeq
Note that the integral over the virtual momenta is convergent. On the one hand, at large momenta the mirror is effectively transparent, so the integral reduces to the integral over the standard empty space modes. On the other hand, at small momenta we can introduce an IR cut-off $p \sim \frac{1}{\Lambda}$ with a clear physical meaning (see Sec.~\ref{sec:two}). Also one can check that this integral is not zero if the motion of the mirror is nonuniform. 

The calculations for the anomalous quantum average~\eqref{eq:k} are essentially the same. Indeed, the leading approximation to the $\kappa_{pq}$ can be expressed in the following form:
\beq \label{kappapq}
\kappa_{pq} = -2 \lambda^2 \int \frac{dk_1dk_2dk_3}{(2\pi)^3} \int_{t_0}^T dt_1 \int_{-\infty}^{+\infty} dx_1 \, g_{k_1,1} g_{k_2,1} g_{k_3,1} \left[ g^*_{p,1} I_q^*(t_1) + g^*_{q,1} I_p^*(t_1) \right]. \eeq
Thus, the final expression for the anomalous quantum average is also proportional to $T^2$:
\beq \kappa_{pq} = -\frac{(\lambda T)^2}{32 \sqrt{|pq|}} \int \frac{dk_1 dk_2 dk_3}{(2\pi)^3 |k_1k_2k_3|} h(p, q, k_1, k_2, k_3) + \mO(\lambda^2 T), \eeq
where the function $h(p, q, k_1, k_2, k_3)$ can be restored after the calculation of integrals~\eqref{kappapq}. This integral is convergent and nonzero due to the same reasons as the integral~\eqref{eq:npq} for $n_{pq}$.

Thus, the DCE with a single nonideal mirror is essentially equal to the case of an ideal mirror considered in~\cite{Alexeev}. The only difference is the presence of a dimensionful parameter $\alpha$ which determines the natural UV scale of the theory. At the same time, we are mainly interested in the secular growth of loop corrections which is essentially an IR effect. Therefore, it is not surprising that such a modification of the theory does not affect the behavior of the loop integrals~\eqref{eq:n} and~\eqref{eq:k}.

\section{Discussion and Conclusion}
\label{sec:discussion}

In this paper we have analyzed the role of nonlinearities in the DCE, i.e. calculated quantum loop corrections to the correlation functions of a self-interacting scalar field on the background of nonuniformly moving mirrors. We have considered the cases of two ideal mirrors and single nonideal mirror. We have shown that in both cases two-loop corrections to the Keldysh propagator~\eqref{eq:K} quadratically grow with the evolution time. This implies that the stress-energy flux~\eqref{eq:T}, energy level density~\eqref{eq:n} and anomalous quantum average~\eqref{eq:k} also receive secularly growing loop corrections. Hence, for large evolution times ($T \sim 1/\lambda \Lambda$) loop corrections are significant even if $\lambda \rightarrow 0$. This indicates a breakdown of the semiclassical approximation. Also this means that for large times particle creation in the DCE should be reconsidered.

We would like to underline several important points concerning our analysis. First of all, once again we emphasize that semiclassical approach to the particle creation cannot be applied to an interacting theory. In fact, the number of created particles in this approach is calculated in terms of the Bogoliubov coefficients~\cite{Birrell}:
\beq \label{eq:tree-N}
\mathcal{N}_m = \langle in | (a_m^{\text{out}})^\dagger a_m^\text{out} | in \rangle = \sum_n | \beta_{nm} |^2, \quad \text{where} \quad \beta_{mn} = - \left( g_m^\text{out}, (g_n^\text{in})^* \right) \eeq
and $( \cdot, \cdot )$ denotes the Klein-Gordon inner product. Here we have introduced the field decomposition in the asymptotic past and future:
\beq \phi(t,x) = \begin{cases} \sum_n \left[ a_n^\text{in} g_n^\text{in}(t,x) + h.c. \right], \quad & \text{when} \quad t \rightarrow -\infty, \\ \sum_n \left[ a_n^\text{out} g_n^\text{out}(t,x) + h.c. \right], \quad & \text{when} \quad t \rightarrow +\infty, \end{cases} \eeq
and defined in- and out-states as $a_n^\text{in} | in \rangle = 0$ and $a_n^\text{out} | out \rangle = 0$. In the notations of Sec.~\ref{sec:SK} quantity $\mathcal{N}_m$ corresponds to the diagonal part\footnote{The relation $a_n^\text{out} = \sum_{m} \left[ \alpha_{nm}^* a_m^\text{in} - \beta_{nm}^* (a_m^\text{in})^\dagger \right]$ straightforwardly implies $\mathcal{N}_m = \sum_{n,k} \beta_{mn} \beta_{mk}^* \left( \delta_{nk} + n_{nk} \right)$, where $n_{nk} = \langle in | (a_n^\text{in})^\dagger a_k^\text{in} | in \rangle$. As we have discussed in the Sec.~\ref{sec:SK}, in the free theory $n_{nk} = 0$, hence, $\mathcal{N}_m = \sum_n |\beta_{nm} |^2$.} of the tree-level energy level density. However, in Secs.~\ref{sec:two} and~\ref{sec:non} we have shown that for sufficiently large evolution times loop corrections to $n_{pq}$ are of the same order as $\delta_{pq}$. Note that these corrections \textit{multiply} the answer~\eqref{eq:tree-N}. Moreover, loop corrections also affect the non-diagonal parts of $n_{pq}$ and anomalous quantum average $\kappa_{pq}$. Hence, the semiclassical estimate~\eqref{eq:tree-N} is incomplete.

We expect the same reasoning to apply to other models of the DCE, including four-dimensional setups. In fact, it is believed that in some approximation the modes of the resonant cavity decouple~\cite{Dodonov:2005, Schutzhold, Dodonov:1990, Dodonov:1995, Dodonov:2020}. In this approximation particle creation is qualitatively described by a quantum harmonic oscillator with time-dependent frequency. However, this qualitative model already posses a kind of secular growth when anharmonic terms are included in the Hamiltonian~\cite{Trunin-2}. This indicates that loop corrections play an important role even in such simple models of the DCE.

It is noteworthy that both the tree-level and loop-level stress-energy fluxes are associated with the violation of the conformal invariance. At the tree-level this invariance is violated by non-trivial boundary conditions~\cite{Davies:1976}. In addition, the secular growth of the loops reflects the conformal non-invariance of the $\lambda \phi^4$ interaction term. This non-invariance directly brings the conformal factor into loop integrals (Sec.~\ref{sec:conformal}).

However, note that loop corrections to the quantum averages and stress-energy flux do not grow with time if trajectories of mirrors are synchronized. Roughly speaking, synchronization forces the mode functions to ``forget'' the periods of nonstationary motion (see the discussion at the end of Sec.~\ref{sec:synch}). The tree-level stress-energy flux on such a background is also zero~\cite{Davies:1976}. We find it remarkable that the absence of particle creation is observed both at the tree-level and in loops.

Finally, we emphasize that the definitive conclusion about the destiny of the vacuum state and the stress-energy flux in the DCE can be made only after the resummation of the leading corrections from all loops. To perform such a resummation one needs to solve a system of the Dyson--Schwinger equations~\cite{Akhmedov, Burda, Polyakov, Landau:vol10}. As a result, the tree-level quantum averages in~\eqref{eq:K} would be replaced by their renormalized values. Furthermore, in some special cases the system of Dyson--Schwinger equations reduce to a Boltzman kinetic equation which suggests a simple physically meaningful solution. Examples of such cases can be found in~\cite{Akhmedov:dS, Pavlenko, Popov, Akhmedov:Et, Akhmedov:Ex}.

Unfortunately, it is still unclear how to perform such a resummation for the DCE. On the one hand, we expect that higher loop corrections to the Keldysh propagator will not be suppressed in the limit $\lambda \rightarrow 0$, $T \rightarrow \infty$, $\lambda T = \const$. In particular, the analysis similar to the Sec.~\ref{sec:conformal} implies that one-loop corrections to the vertexes quadratically grow in this limit. On the other hand, the quadratic growth of the two-loop correction to the propagator is itself unconventional. Due to these reasons Dyson--Schwinger equations of the theory~\eqref{eq:S} do not reduce to the kinetic equation\footnote{This also means that the interpretation of $n_{pq}$ and $\kappa_{pq}$ as the energy level density and anomalous quantum average may be misleading. It is safer to find the exact Keldysh propagator~\eqref{eq:K} first and analyze its structure afterwards.}. Therefore, one needs to develop a completely new method to solve these equations. We hope that the analysis of this paper will help to deduce such a method.

\section*{Acknowledgements}

We would like to thank Emil Akhmedov for proposing to us this problem, valuable discussions and proof-reading of the text. Also we would like to thank Sergey Alexeev, Kirill Gubarev and Gheorghe Sorin Paraoanu for useful comments. LAA would like to thank Hermann Nicolai and Stefan Theisen for the hospitality at the Albert Einstein Institute, Golm, where the work on this project was partly done. The work of DAT was supported by the Russian Ministry of education and science and by the grant from the Foundation for the Advancement of Theoretical Physics and Mathematics ``BASIS''.

\appendix

\section{False secular growth in the stationary theory}
\label{sec:artifacts}

Let us show that a stationary theory may possess a fake ``secular growth'' at evolution times much less than the natural IR cut-off, although in the limit $T \rightarrow \infty$ both $n_{pq} \rightarrow 0$ and $\kappa_{pq} \rightarrow 0$. Consider two ideal mirrors located at points $x = 0$ and $x = \Lambda$. This setup corresponds to the following equation of motion for a massless scalar field:
\beq \pd_\mu \pd^\mu \phi = 0, \quad \phi(t, 0) = \phi(t, \Lambda) = 0, \eeq
which imply the following mode decomposition:
\beq \label{eq:false-1}
\phi(t,x) = \sum_{n = 1}^\infty \left[ a_n g_n(t,x) + a_n^\dagger g_n^*(t,x) \right], \quad g_n(t,x) = \frac{e^{-i \frac{\pi n}{\Lambda} t}}{\sqrt{\pi n}} \sin \frac{\pi n x}{\Lambda}, \eeq
where $\left[ a_m, a_n^\dagger \right] = \delta_{mn}$. Now let us remind that the two-loop correction to $n_{pq}$ reduces to the product of two integrals (we can set $t_0 = 0$ due to the time translation invariance):
\beq n_{pq}(T) = 2 \lambda^2 \sum_{m,n,k} I_{p,m,n,k}(T) I_{q,m,n,k}^*(T), \quad I_{p,m,n,k}(T) = \int_0^T dt \int_0^\Lambda dx g_p(t,x) g_m(t,x) g_n(t,x) g_k(t,x) \eeq
Substituting the modes~\eqref{eq:false-1} into these integrals we obtain:
\beq \begin{aligned}
I_{p,m,n,k}(T) &= \int_0^T dt \int_0^\Lambda dx \frac{e^{-\frac{i \pi t}{\Lambda}(p+m+n+k)}}{8 \pi^2 \sqrt{p m n k}} \sum_{\sigma_m, \sigma_n, \sigma_k = \pm 1} \sigma_m \sigma_n \sigma_k \cos \left[ \frac{\pi x}{\Lambda} (p + \sigma_m m + \sigma_n n + \sigma_k k) \right] = \\ &= \frac{1 - e^{-\frac{i \pi T}{\Lambda}(p+m+n+k)}}{\frac{i \pi}{\Lambda}(p+m+n+k)} \frac{1}{8 \pi^2 \sqrt{p m n k}} \sum_{\sigma_m, \sigma_n, \sigma_k = \pm 1} \sigma_m \sigma_n \sigma_k \Lambda \delta_{p + \sigma_m m + \sigma_n n + \sigma_k k, 0}.
\end{aligned} \eeq
Now it is straightforward to see that at small times, $T \ll \Lambda$, the integral linearly grows with time, $I_{p,m,n,k}(T) \sim \Lambda T$. Hence, at such times the energy level density also grows secularly, $n_{pq}(T) \sim (\lambda \Lambda T)^2$. However, at large times, $T \gg \Lambda$, the time-dependent part reduces to the Dirac delta-function whose argument is never zero, $\frac{1 - e^{-\frac{i \pi T}{\Lambda}(p+m+n+k)}}{\frac{i \pi}{\Lambda}(p+m+n+k)} \rightarrow \delta\left( \frac{\pi(p + m + n + k)}{\Lambda} \right) = 0$. Therefore, in this limit the correction to the energy level density is approximately zero, $n_{pq} \rightarrow 0$, as it should be. Similarly one can show that $\kappa_{pq} \sim -(\lambda \Lambda T)^2$ for $T \ll \Lambda$ and $\kappa_{pq} \rightarrow 0$ for $T \gg \Lambda$.

This behavior of the quantum averages has a clear physical interpretation. At large evolution times the energy conservation and momentum conservation\footnote{Note that in this case the ``momentum conservation'' means the momentum conservation in the Brillouin zone, i.e. the total momentum can change by $\frac{2 \pi n}{\Lambda}$, $n \in \mathbb{Z}$.} laws exclude any energy exchange; this means that the collision integral is zero and quantum averages cannot receive nonzero loop corrections. However, at small evolution times the energy conservation law may be violated. Hence, usually forbidden processes are allowed, so quantum averages may temporarily grow. Since this growth does not persist for long evolution times, it reflects mere vacuum fluctuations rather than the change in the state of the system. Thus, we need to distinguish between this effect and the true secular growth.

\section{Mode functions for the case of two kicks}
\label{sec:expl-modes}

In this appendix we present the explicit expressions for the functions $G[t,\delta] = G\left[ (1 + \beta) t + \delta \right]$, $F[t,\delta] = F\left[ (1 - \beta) t - \delta \right]$ on the background~\eqref{eq:sim-tr}. Since these functions increase by $2$ when $t$ increases by $\tau = \frac{2 \Lambda}{1 - \beta^2}$, it is convenient to introduce the number  $k$:
\beq k \equiv \left\lceil \frac{t - \Lambda}{\tau} \right\rceil. \eeq
Then it is straightforward to show that $G\left[ t>0, \delta \right]$ and $F\left[ t>0, \delta \right]$ are given by the following piecewise linear functions (we assume that $0 < \beta < 1$; the expressions for the negative $\beta$ can be obtained in the similar way):
\small \begin{align}
\label{eq:G-sim-expl} G(t, \delta) &= \begin{cases} \frac{(1 - \beta) t}{\Lambda} + \frac{1 - \beta}{1 + \beta} \frac{\delta}{\Lambda} + \frac{2 \beta k}{1 + \beta}, \quad &\text{if} \quad \begin{cases}  k - \frac{1 + \beta^2}{2} < \frac{t}{\tau} < k - \frac{1 - \beta}{2}, \quad 0 < \delta < \Lambda, \\ k - \frac{1 - \beta}{2} < \frac{t}{\tau} < k, \quad 0 < \frac{\delta}{\Lambda} < \frac{2}{1 - \beta} \left( k - \frac{t}{\tau} \right), \end{cases} \\
\frac{(1 + \beta) t}{\Lambda} + \frac{\delta}{\Lambda} - \frac{2 \beta k}{1 - \beta}, \quad &\text{if} \quad \begin{cases} k - \frac{1 - \beta}{2} < \frac{t}{\tau} < k, \quad \frac{2}{1 - \beta} \left( k - \frac{t}{\tau} \right) < \frac{\delta}{\Lambda} < 1, \\ k < \frac{t}{\tau} < k + \frac{1 - \beta}{2}, \quad 0 < \frac{\delta}{\Lambda} < 1 - \frac{2}{1 - \beta} \left( \frac{t}{\tau} - k \right), \end{cases} \\
\frac{(1 - \beta) t}{\Lambda} + \frac{1 - \beta}{1 + \beta} \frac{\delta}{\Lambda} + \frac{2 \beta (k + 1)}{1 + \beta}, \quad &\text{if} \quad \begin{cases}  k < \frac{t}{\tau} < k + \frac{1 - \beta}{2}, \quad 1 - \frac{2}{1 - \beta} \left( \frac{t}{\tau} - k \right) < \frac{\delta}{\Lambda} < 1, \\ k + \frac{1 - \beta}{2} < \frac{t}{\tau} < k + \frac{1 - \beta^2}{2}, \quad 0 < \frac{\delta}{\Lambda} < 1, \end{cases} \end{cases} \\
\label{eq:F-sim-expl} F(t, \delta) &= \begin{cases} \frac{(1 - \beta) t}{\Lambda} - \frac{\delta}{\Lambda} + \frac{2 \beta (k - 1)}{1 + \beta}, \quad &\text{if} \quad k - \frac{1 + \beta^2}{2} < \frac{t}{\tau} < k - \frac{1 - \beta}{2}, \quad \frac{2 (t/\tau - k + 1)}{1 + \beta} < \frac{\delta}{\Lambda} < 1, \\
\frac{(1 + \beta) t}{\Lambda} - \frac{1 + \beta}{1 - \beta} \frac{\delta}{\Lambda} + \frac{2 \beta k}{1 - \beta}, \quad &\text{if} \quad \begin{cases} k - \frac{1 + \beta^2}{2} < \frac{t}{\tau} < k - \frac{1 - \beta}{2}, \quad 1 < \frac{\delta}{\Lambda} - \frac{2 (t/\tau - k)}{1 + \beta} < \frac{2}{1 + \beta}, \\ k - \frac{1 - \beta}{2} < \frac{t}{\tau} < k, \quad 1 - \frac{2}{1 - \beta} \left( k - \frac{t}{\tau} \right) < \frac{\delta}{\Lambda} < 1, \end{cases} \\
\frac{(1 - \beta) t}{\Lambda} - \frac{\delta}{\Lambda} + \frac{2 \beta k}{1 + \beta}, \quad &\text{if} \quad \begin{cases} k - \frac{1 + \beta^2}{2} < \frac{t}{\tau} < k - \frac{1 - \beta}{2}, \quad 0 < \frac{\delta}{\Lambda} < 1 - \frac{2 (k - t/\tau)}{1 + \beta}, \\ k - \frac{1 - \beta}{2} < \frac{t}{\tau} < k, \quad 0 < \frac{\delta}{\Lambda} < 1 - \frac{2}{1 - \beta} \left( k - \frac{t}{\tau} \right), \\ k < \frac{t}{\tau} < k + \frac{1 - \beta^2}{2}, \quad \frac{2}{1 + \beta} \left( \frac{t}{\tau} - k \right) < \frac{\delta}{\Lambda} < 1, \end{cases} \\
\frac{(1 + \beta) t}{\Lambda} - \frac{1 + \beta}{1 - \beta} \frac{\delta}{\Lambda} + \frac{2 \beta (k + 1)}{1 - \beta}, \quad &\text{if} \quad \begin{cases} k < t < k + \frac{1 - \beta}{2}, \quad 0 < \frac{\delta}{\Lambda} < \frac{2}{1 + \beta} \left( \frac{t}{\tau} - k \right), \\ \frac{1 - \beta}{2} < \frac{t}{\tau} - k < \frac{1 - \beta^2}{2}, \quad 1 < \frac{\delta}{\Lambda} + \frac{2 (k + 1 - t/\tau)}{1 + \beta} < \frac{2}{1 + \beta}, \end{cases} \\
\frac{(1 - \beta) t}{\Lambda} - \frac{\delta}{\Lambda} + \frac{2 \beta (k + 1)}{1 + \beta}, \quad &\text{if} \quad k + \frac{1 - \beta}{2} < \frac{t}{\tau} < k + \frac{1 - \beta^2}{2}, \quad 0 < \frac{\delta}{\Lambda} < 1 - \frac{2 (k + 1 - t/\tau)}{1 + \beta}. \end{cases}
\end{align} \normalsize
Graphs of these functions for a fixed $\delta$ look like a periodically growing saw. Namely, they are glued from the alternating slow-growing (with the slope $\frac{1 - \beta}{\Lambda}$) and fast-growing (with the slope $\frac{1 + \beta}{\Lambda}$) parts. For large $\beta$ the teeth of the saw are very sharp; on the contrary, for small $\beta$ graphs are almost smooth. The change of $\delta$ simply shifts the graphs up or down.

The mode functions are given by the formula~\eqref{eq:mode-moore} as usual:
\beq g_n(t, \delta) = \frac{i}{\sqrt{4 \pi n}} \left[ e^{-i \pi n G(t,\delta)} - e^{-i \pi n F(t,\delta)} \right]. \eeq
At large $\beta$ and fixed $t$ these functions possess the following behavior. At the largest part of the interval (of the length $(1 - \beta) \Lambda$) they look like a slowly oscillating exponential function $\sim e^{-i \pi n \frac{\delta}{\Lambda} + \varphi_1}$, where the phase $\varphi_1$ does not depend on $\delta$. At the remaining part they oscillate much more rapidly, $\sim e^{-i \pi n \frac{1 + \beta}{1 - \beta} \frac{\delta}{\Lambda} + \varphi_2}$. In this region functions $G$ and $F$ increase by roughly $1$, so the difference of phases to the left and to the right from the fast-oscillation region is roughly $e^{-i \pi n}$. In other words, in even modes left and right slowly-oscillating exponents simply continue each other, whereas in odd modes these exponents change the sign at the gluing point.

\section{Loop corrections}
\label{sec:loop_cor}

The full result for the function $\sum_{\{p\}}J^{\{p\}}_{u,v}$ is presented in this section. Let us introduce two auxiliary integrals:
\begin{align}
&\begin{aligned} Int_1&=[\theta(-p)\theta(k_1)\theta(k_2)\theta(k_3)+\theta(p)\theta(-k_1)\theta(-k_2)\theta(-k_3)] \times\\
&\times \bigg [ R_{|p|}T_{|k_3|}T_{|k_2|}T_{|k_1|}+T_{|p|}R_{|k_3|}R_{|k_2|}R_{|k_1|} -\\
&-\frac{s}{s+\frac{2|p|}{1+c\beta}} \left (R^{c}_{|p|}T^{-c}_{|k_3|}T^{-c}_{|k_2|}T^{-c}_{|k_1|} + T^{c}_{|p|}R^{-c}_{|k_3|}R^{-c}_{|k_2|}R^{-c}_{|k_1|}[D_\beta^{-c}]^2\right )e^{-i(|k_1|+|k_2|+|k_3|)(T-cx(T))-i|p|(T+cx(T))} + \\
&+\int_0^{T-cx(T)} d[-isu] R^{c}_{|p|}(u)T^{-c}_{|k_3|}(u)T^{-c}_{|k_2|}(u)T^{-c}_{|k_1|}(u)e^{-i(|k_1|+|k_2|+|k_3|)u-i|p|v(u)}+ \\
&+\int_0^{T+cx(T)} d[-isv]T^{c}_{|p|}(v)R^{-c}_{|k_3|}(v)R^{-c}_{|k_2|}(v)R^{-c}_{|k_1|}(v) e^{-i(|k_1|+|k_2|+|k_3|)u(v)-i|p|v} \bigg ], \end{aligned} \\
&\begin{aligned} Int_2&=[\theta(p)\theta(k_1)\theta(-k_2)\theta(-k_3)+\theta(-p)\theta(-k_1)\theta(k_2)\theta(k_3)] \times \\
&\times \bigg [ R_{|p|}T_{|k_3|}T_{|k_2|}R_{|k_1|}+T_{|p|}R_{|k_3|}R_{|k_2|}T_{|k_1|} -\\
&-\frac{s}{s+\frac{2(|p|+|k_1|)}{1+c\beta}} \left (R^{c}_{|p|}T^{-c}_{|k_3|}T^{-c}_{|k_2|}R^{c}_{|k_1|} + T^{c}_{|p|}R^{-c}_{|k_3|}R^{-c}_{|k_2|}T^{c}_{|k_1|}[D_\beta^{-c}]^2\right )e^{-i(|k_2|+|k_3|)(T-cx(T))-i(|p|+|k_1|)(T+cx(T))} + \\
&+\int_0^{T-cx(T)} d[-isu] R^{c}_{|p|}(u)T^{-c}_{|k_3|}(u)T^{-c}_{|k_2|}(u)R^{c}_{|k_1|}(u)e^{-i(|k_2|+|k_3|)u-i(|k_1|+|p|)v(u)}+ \\
&+\int_0^{T+cx(T)} d[-isv]T^{c}_{|p|}(v)R^{-c}_{|k_3|}(v)R^{-c}_{|k_2|}(v)T^{-c}_{|k_1|}(v) e^{-i(|k_2|+|k_3|)u(v)-i(|k_1|+|p|)v} \bigg ]. \end{aligned}
\end{align}
where $c = \sgn(p)$, $s = |p| + |k_1| + |k_2| + |k_3|$, $v(u) = t_u + x(t_u)$, $u(v) = t_v - x(t_v)$. Then the final result for the function $J$ can be written down in terms of the introduced integrals:
\beq \begin{aligned} \sum_{\{p\}}J^{\{p\}}_{u,v}
&= Int_1 + Int_1(p\leftrightarrow k_1)+Int_1(p\leftrightarrow k_2)+Int_1(p\leftrightarrow k_3) + \\
&+ Int_2 + Int_2(k_1\leftrightarrow k_2) + Int_2(k_1\leftrightarrow k_3).
\end{aligned} \eeq

\end{document}